\documentclass[prb,aps,english,twocolumn,notitlepage,superscriptaddress]{revtex4-2}

\usepackage{graphicx}% Include figure files
\usepackage{dcolumn}% Align table columns on decimal point
\usepackage{rotating}
\usepackage{lipsum}
\usepackage{xcolor}
\usepackage{bm}% bold math
\usepackage[T1]{fontenc}  
\usepackage{multirow,amsmath,amssymb}
 \usepackage{babel} 
 \usepackage{txfonts}
 \usepackage{epsfig,epsf,psfrag} 
\usepackage{graphicx} 
\usepackage{pslatex}
\usepackage{epic,eepic} 
\usepackage{color,pstcol}
\usepackage{pstricks} 
\usepackage{fancyhdr} 
\usepackage{tabularx}
\usepackage{physics}
\usepackage{url}
\usepackage{listings}
\usepackage{color} 
\usepackage{caption}
\usepackage{ulem}
\usepackage{balance}
\usepackage{subcaption}
\usepackage{float}
\usepackage[utf8]{inputenc}
\usepackage{comment}

\usepackage{titlesec}
\titleclass{\subsubsubsection}{straight}[\subsubsection]
\newcounter{subsubsubsection}[subsubsection]
\renewcommand\thesubsubsubsection{\thesubsubsection.\arabic{subsubsubsection}}

\titleformat{\subsubsubsection}
  [block] % ← Use 'block' for center alignment
  {\centering\normalfont\normalsize\bfseries} % ← centering added here
  {\thesubsubsubsection} % number format
  {1em} % spacing between number and title
  {}

\titlespacing*{\subsubsubsection}
  {0pt}{3.25ex plus 1ex minus .2ex}{1.5ex plus .2ex}

\usepackage{hyperref}
\hypersetup{
 colorlinks=true,
 linkcolor=blue,
 anchorcolor = blue,
 citecolor = blue,
 filecolor = blue,
 urlcolor = blue
}
\usepackage{dsfont}

\graphicspath{{Images/}}
\def \be {\begin{equation}} 
\def \ee {\end{equation}}

\begin{document}

\preprint{APS/123-QED}
\title{
Temperature-Bias Noise and Quantum Shot Noise as Probes of Pairing Symmetry in Iron Pnictides
}% Force line breaks with \\
% Detection of pairing symmetry in Iron Pnictide superconductor via quantum shot noise, thermovoltage and excess current

\author{Sachiraj Mishra}
\email{sachiraj29mishra@gmail.com}
\author{A Rajmohan Dora}
\email{a.rajmohandora@outlook.com}
\author{Colin Benjamin}
\email{colin.nano@gmail.com}
\affiliation{School of Physical Sciences, National Institute of Science Education and Research, HBNI, Jatni-752050, India}
\affiliation{Homi Bhabha National Institute, Training School Complex, AnushaktiNagar, Mumbai, 400094, India }

% \title{Detection of pairing symmetry in Iron Pnictide superconductor via $\Delta_T$ noise
% }

\begin{abstract}
    {

Quantum noise has long served as a powerful probe of quantum transport in mesoscopic junctions. Recently, temperature-driven noise, or $\Delta_T$ noise, has attracted growing interest due to its presence even in the absence of average charge current. In this work, we investigate a normal metal–insulator–iron-pnictide junction and demonstrate how zero temperature quantum shot noise, finite temperature quantum noise and $\
\Delta_T$ noise can discriminate between $S_{++}$ and $S_{+-}$ pairing symmetries, which are relevant to iron-based superconductors. We introduce $\Delta_T$ noise as a novel probe for distinguishing between the two pairing symmetries. In contrast to conductance, which exhibits a single peak for both $S_{++}$ and $S_{+-}$ states with only a difference in magnitude, the $\Delta_T$ noise reveals qualitatively distinct features: a twin-peak structure for the $S_{++}$ pairing symmetry and a single-peak profile for the $S_{+-}$ state. A similar symmetry-dependent contrast is observed in both zero temperature quantum shot noise and finite temperature quantum 
noise, where the $S_{++}$ state consistently exhibits a twin-peak structure, while the $S_{+-}$ state shows a single-peak response. Our results demonstrate that noise-based measurements form a mutually reinforcing set of probes that enables reliable identification of superconducting gap symmetry in Iron Pnictide superconductors.

    }
\end{abstract}

\maketitle

%###############################################################
%###############################################################
%###############################################################
%###############################################################
\section{Introduction}

% In Normal metal-Insulator-Iron Pnictide superconductor(N-I-IP) junction, the current becomes zero only at applied voltage bias $V=0$.

The detection and identification of the pairing symmetry in iron-pnictide superconductors remains a central and unresolved issue in the field of unconventional superconductivity. These materials are known for their multiband nature, and understanding the symmetry of the superconducting order parameter is essential for uncovering the pairing mechanism. In this study, we propose a method to distinguish between the $S_{++}$ and $S_{+-}$ pairing symmetries using thermally-induced current fluctuations, specifically through the analysis of quantum shot noise, and $\Delta_T$ noise in a normal metal-insulator-Iron Pnictide (N-I-IP) superconductor junction.

The nature of the relative phase between these gap parameters determines the symmetry of the superconducting order parameter. If the superconducting gaps on different bands have the same phase, the pairing symmetry is classified as $S_{++}$. On the other hand, if the gap functions have opposite phases, i.e., $\phi_{2} = \phi_{1}\pm \pi$, the symmetry is referred to as $S_{+-}$\cite{PhysRevB.80.144507}. Understanding the superconducting pairing symmetry in iron-pnictide superconductors has been a topic of extensive theoretical and experimental research. In particular, distinguishing between the $S_{++}$ and $S_{+-}$ pairing symmetries is crucial for unveiling the underlying pairing mechanism. Experimental techniques such as angle-resolved photoemission spectroscopy (ARPES)~\cite{Nakayama2009EPL}, neutron scattering, and tunnelling spectroscopy, along with theoretical models based on multiband and spin fluctuation-mediated pairing, have provided significant insights into this issue~\cite{PhysRevLett.101.057003, Hirschfeld_2011, Chubukov, PhysRevLett.104.157001, Christianson2008}. Previously, the pairing symmetry of Iron Pnictides has been studied theoretically using zero temperature differential shot noise\cite{Benjamin2020}, conductance and Josephson super-current \cite{PhysRevB.80.144507}. 

Iron-pnictide superconductors typically possess a complex electronic structure featuring multiple Fermi surfaces. In the minimal two-band model\cite{PhysRevB.77.220503}, which captures the essential physics, the system exhibits two electron pockets and two hole pockets, as revealed by tight-binding calculations \cite{PhysRevB.77.220503} and angle-resolved photoemission spectroscopy (ARPES). These multiple Fermi pockets naturally give rise to multiple superconducting gaps, denoted as $\Delta_1$ and $\Delta_2$, associated with different bands. So far, there has been no clear consensus on the pairing symmetries in Iron Pnictide superconductors~\cite{Chubukov}; our work aims to provide a method to resolve this ambiguity. At present, no complete microscopic theory exists for unconventional superconductors, and identifying their pairing symmetries is a crucial step toward developing a non-BCS theoretical framework for Iron Pnictide superconductors. The multiband character and metallic nature of Iron Pnictides offer a distinct and potentially more tractable platform for investigating unconventional superconductivity, in contrast to the strongly correlated, single-band cuprates~\cite{Norman2008}.

% In a Josephson junction, $s_{+-}$ pairing causes competing band contributions that can produce a $0\text{–}\pi$ \text{ transition in the current–phase relation, whereas } $s_{++}$ \text{ gives only a } 0\text{-junction.} Although phase-sensitive Josephson measurements have been proposed as a promising probe of the $S_{+-}$ pairing symmetry \cite{PhysRevB.80.144507}, such experiments require superconductor–superconductor junctions and precise phase control, making their realization experimentally challenging. In contrast, noise spectroscopy and thermoelectric effects in normal metal–Iron Pnictide superconductor junctions can provide complementary signatures of the pairing symmetry in a simpler transport geometry. 

Beyond conductance-based approaches, Josephson junctions have been widely used as phase-sensitive probes to investigate the pairing symmetry in iron pnictides. In particular, the sign-changing $s_{\pm}$ order parameter leads to competing contributions from different bands, resulting in characteristic effects such as $0\text{--}\pi$ transitions, phase shifts, and the emergence of higher harmonics in the current--phase relation \cite{PhysRevB.80.144507}. Additional theoretical studies of hybrid junctions further predict that interband coupling and spin polarization can induce oscillatory behavior and $0\text{--}\pi$ transitions that are absent in the conventional $s_{++}$ case, providing potential distinguishing signatures \cite{Liu2014Interplay}. Experimentally, the Josephson current has also been shown to exhibit strong sensitivity to Fermi surface matching and interface engineering, such as the anomalous enhancement of the $I_c R_N$ product upon insertion of a normal metal interlayer, highlighting the role of band-dependent tunneling in $s_{\pm}$ systems \cite{PhysRevB.103.214507}. Furthermore, detailed measurements of the current--phase relation and the temperature dependence of the critical current have been found to be consistent with multiband $s_{\pm}$ models under appropriate conditions \cite{Stepanov2021}. 

However, while these studies provide strong phase-sensitive evidence for a sign-changing $s_{\pm}$ order parameter, their interpretation relies on the specific junction configuration and modeling of multiband effects \cite{PhysRevB.80.144507,PhysRevB.103.214507}. In particular, the Josephson response reflects the combined contributions of multiple bands, whose relative weights depend on factors such as interface transparency, Fermi surface matching, and interlayer properties. Consequently, features such as $0\text{--}\pi$ transitions and modifications of the current--phase relation can be influenced by junction-specific conditions and require careful theoretical analysis for quantitative interpretation. Additionally, experimental extraction of the current--phase relation or phase shifts often involves indirect techniques, such as microwave irradiation or SQUID measurements, which can further complicate the analysis. These considerations do not diminish the significance of Josephson probes, but indicate that complementary approaches can be valuable for obtaining a more complete and robust identification of the pairing symmetry.

% However, despite these advances, these signatures do not by themselves constitute a universally unambiguous determination of the pairing symmetry. The reported Josephson effects provide strong phase-sensitive evidence consistent with a sign-changing $s_{\pm}$ order parameter, but are typically interpreted within specific junction conditions and models \cite{PhysRevB.80.144507,PhysRevB.103.214507}. 

Although conductance measurements and Josephson junctions provide valuable information about quasiparticle transport, they are often insufficient for unambiguously distinguishing between different pairing symmetries in multiband superconductors. In particular, both $S_{++}$ and $S_{+-}$ states typically exhibit similar conductance spectra, characterized by a single peak structure that differs primarily in magnitude rather than in qualitative features. As a result, conductance alone does not offer a definitive fingerprint of the underlying superconducting order parameter. This limitation motivates the exploration of alternative transport probes that are more sensitive to interband coupling and electron-hole asymmetry, and hence better suited for identifying the pairing symmetry. A recent thermoelectric study also distinguishes different pairing symmetries in iron-pnictide superconductors using thermoelectric signatures \cite{PhysRevB.108.L100511}. However, use of finite temperature quantum noise and $\Delta_T$ noise has never been enforced to detect pairing symmetry of Iron Pnictide superconductor.

$\Delta_T$ noise in mesoscopic junctions refers to quantum shot noise generated under a finite temperature gradient, when the average charge current transported is zero. In this paper, we show that it can be a very effective tool to probe the pairing symmetry of Iron Pnictide superconductors, see Ref.~\cite{Lumbroso2018}, as it originates purely from the temperature gradient. $\Delta_T$ noise has attracted considerable attention in recent years from both 
theoretical and experimental perspectives. A number of studies have explored 
$\Delta_T$ noise in a wide variety of physical systems, including atomic-scale 
molecular junctions~\cite{Lumbroso2018}, quantum circuits~\cite{Sivre2019}, and 
metallic tunnel junctions~\cite{PhysRevLett.125.106801}, among others%
~\cite{Zhitlukhina2020, Eriksson2021, Popoff2022, Tesser2023, AndreevDTnoise}. At finite temperature, the total quantum noise contains contributions 
from both shot noise and thermal fluctuations\cite{Blanter2000}, leading to a richer 
structure than in the zero-temperature limit. The interplay between 
thermal broadening and bias-driven quasiparticle transport modifies 
the magnitude and position of the characteristic features in the noise 
spectra. Consequently, finite-temperature quantum noise provides an 
additional and experimentally relevant probe of pairing symmetry in 
multiband superconductors.

In this work, we have employed a quantum transport-based approach to probe the phase structure of Iron Pnictides. Further analysing the behaviour of zero temperature quantum shot noise, finite temperature quantum noise and $\Delta_T$ noise across the junction, we demonstrate that these observables too exhibit markedly different features depending on whether the superconductor exhibits $S_{++}$ or $S_{+-}$ symmetry. Our findings suggest that $\Delta_T$ noise, zero temperature quantum shot noise, finite temperature quantum noise can serve as sensitive and experimentally accessible probe for detecting the underlying pairing symmetry in iron-pnictide superconductors.

% Theory starts

{The structure of this paper is as follows: in Sec.~\ref {Formalism}(A-C) we outline the theoretical framework employed in our study, and focus on the calculation of current in a N-I-IP junction setup using the Landauer--B\"uttiker scattering approach. In Sec.~\ref{Formalism}(D-E), we present the general theory of quantum noise in N-I-IP junction. In Sec.~\ref {results}, we present our results, which involve zero temperature quantum shot noise, finite temperature quantum noise and $\Delta_T$ noise. In Sec. ~\ref{analysis}, we analyse via a table the differences between the $S_{++}$ and $S_{+-}$ pairing symmetries. In Appendix~\ref {s matrix}, we discuss the derivation of the scattering matrix associated with this N-I-IP junction. In Appendix \ref{quantum noise appendix}, we derive the finite-temperature quantum noise in the N-I-IP setup. 
} 

%###############################################################
%###############################################################
%###############################################################
%###############################################################
\section{Theory}
\label{Formalism}

%@@@@@@@@@@@@@@@@@@@@@@@@@@@@@@@@@@@@@@@
\subsection{Hamiltonian and the pairing symmetry of Iron Pnictides}
\label{sec2a}
%@@@@@@@@@@@@@@@@@@@@@@@@@@@@@@@@@@@@@@@

%*****************************************************************************************

%*****************************************************************************************

{A schematic of the Normal metal-Insulator-Iron Pnictide superconductor (N-I-IP) junction is shown in Fig. \ref{setup1}, where the insulator is modeled by the potential $V \delta(x)$ at $x = 0$, and $V$ represents the strength of barrier potential at the interface. The Hamiltonian for the N-I-IP junction:~\cite{PhysRevB.80.144507}}
{
\begin{equation}
    \mathcal{H} = \begin{pmatrix}
        H_{01} & \Delta_1\theta(x) &{\tilde{\alpha}}_0\delta(x) & 0 \\
        \Delta_1^*\theta(x) & -H_{01} & 0 & -{\tilde{\alpha}}_0\delta(x) \\
        {\tilde{\alpha}}_0\delta(x) & 0 & H_{02} & \Delta_2\theta(x) \\
        0 & -{\tilde{\alpha}}_0\delta(x) & \Delta_2^*\theta(x) & -H_{02}
    \end{pmatrix},
\end{equation}}
{where {$H_{01} = \frac{\hbar^2 k_1^2}{2m*} + V \delta(x) - E_F$} is associated with band 1 of superconductor, while {$H_{02} = \frac{\hbar^2 k_2^2}{2m*} + V\delta(x) - E_F$} is associated with band 2 of superconductor. Here, $m^*$ is the effective mass of the electron, and $E_F$ is the Fermi energy of the system. $\Delta_1$ and $\Delta_2$ are the superconducting gaps associated with band 1 and band 2, respectively. The interband coupling strength between two bands is ${\tilde{\alpha}}_0$~\cite{PhysRevB.80.144507}. $\theta(x)$ is the Heaviside theta function and $\delta(x)$ is the delta function.
% The iron-pnictide superconductor is modelled as a two-band system with gaps $\Delta_1$ and $\Delta_2 e^{-i\phi}$, where $\phi$ is the relative phase between bands. 

\begin{figure}[H]
    \centering
    \includegraphics[width=0.8\linewidth]{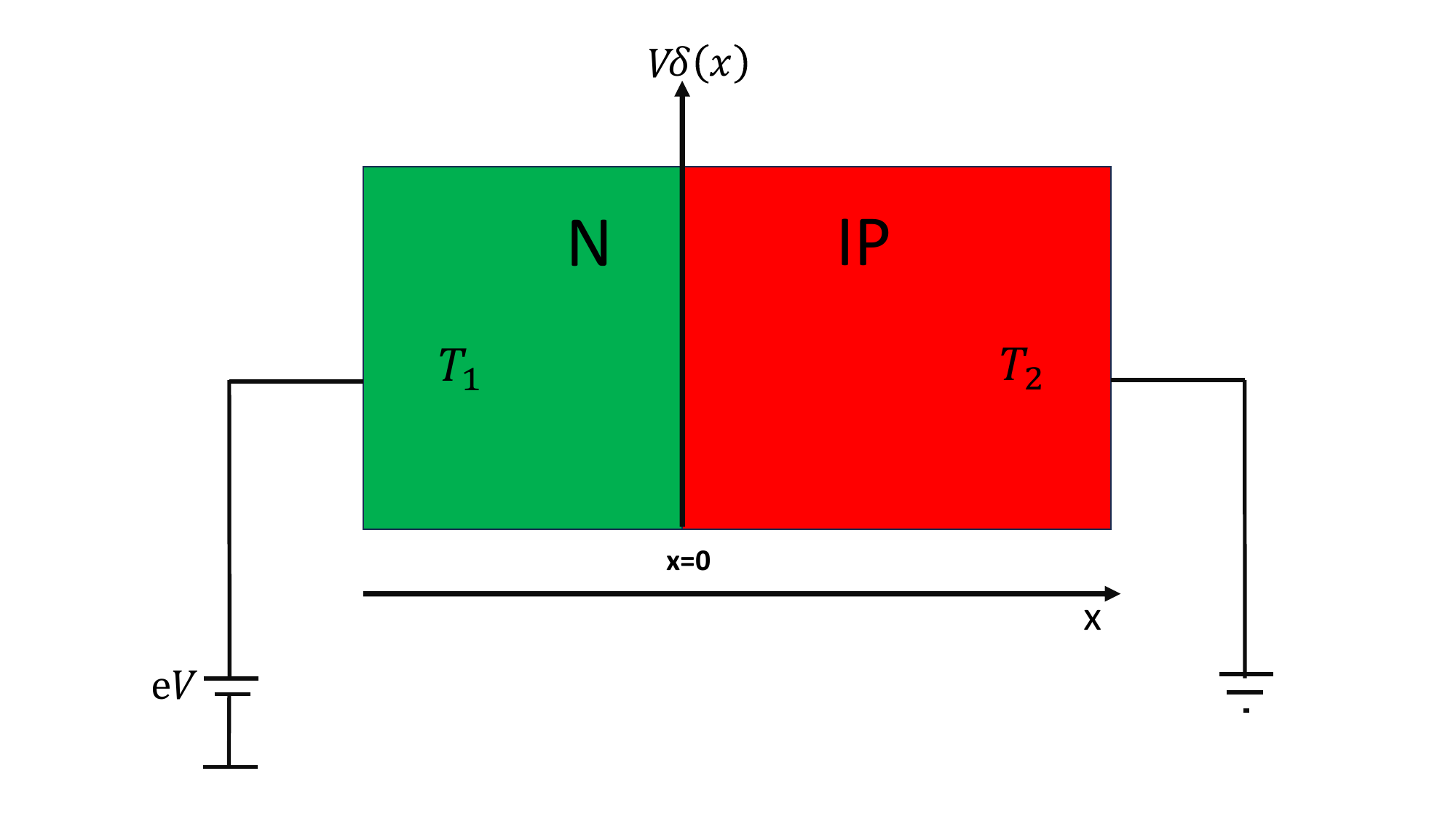}
    \caption{Schematic of the N-I-IP junction: Normal metal is at temperature $T_1$, Iron-pnictide superconductor is at temperature $T_2$, with voltage $V$ applied to normal metal while Iron-pnictide superconductor is grounded.}
    \label{setup1}
\end{figure}

The wave functions for an electron incident from the normal metal in band $1$ are given below.
\begin{widetext}

\begin{subequations}\label{eq:psi}
\begin{align}
\psi_{N}(x) &=
\left(e^{ik_1 x} + r^{ee}_{11n} e^{-ik_1 x}\right)\phi^{N}_1
+ r^{he}_{11n} e^{ik_2 x}\phi^{N}_2
+ r^{ee}_{12n} e^{-ik_1 x}\phi^{N}_3
+ r^{he}_{12n} e^{ik_2 x}\phi^{N}_4, 
&& x<0, \label{eq:psi-2a} \\[6pt]
\psi_{IP}(x) &=
t^{ee}_{11n} e^{iq_{1e} x}\phi^{S}_1
+ t^{he}_{11n} e^{-iq_{1h} x}\phi^{S}_2
+ t^{ee}_{12n} e^{iq_{2e} x}\phi^{S}_3
+ t^{he}_{12n} e^{-iq_{2h} x}\phi^{S}_4, 
&& x>0, \label{eq:psi-2b}
\end{align}
\label{wavefunction}
\end{subequations}

\[
\text{with } \phi_1^N = \begin{pmatrix} 1 \\ 0 \\ 0 \\ 0 \end{pmatrix}, \quad
\phi_2^N = \begin{pmatrix} 0 \\ 1 \\ 0 \\ 0 \end{pmatrix}, \quad
\phi_3^N = \begin{pmatrix} 0 \\ 0 \\ 1 \\ 0 \end{pmatrix}, \quad
\phi_4^N = \begin{pmatrix} 0 \\ 0 \\ 0 \\ 1 \end{pmatrix}, \quad
\phi_1^S = \begin{pmatrix} u_1 \\ v_1 \\ 0 \\ 0 \end{pmatrix}, \quad
\phi_2^S = \begin{pmatrix} v_1 \\ u_1 \\ 0 \\ 0 \end{pmatrix} , \quad
\phi_3^S = \begin{pmatrix} 0 \\ 0 \\ u_2 \\ v_2 e^{-i\phi} \end{pmatrix}, \quad
\phi_4^S =  \begin{pmatrix} 0 \\ 0 \\ v_2 e^{i\phi} \\ u_2 \end{pmatrix}
\]
\end{widetext}

In $\psi_{N}$, Eq.~\eqref{eq:psi-2a} the amplitude of the incident electron is unity, $r^{ee}_{11n}$ is the normal reflection from band 1, $r^{ee}_{12n}$ is the normal reflection from band 2, $r^{he}_{11n}$ represents the Andreev reflection from band 1, and $r^{he}_{12n}$ represents the Andreev reflection from band 2. In $\psi_{IP}$, $t^{ee}_{11n}$ is the amplitude of transmission as electron-like quasiparticle from band 1, $t^{ee}_{12n}$ is the amplitude of transmission as electron-like quasiparticle from band 2, $t^{he}_{11n}$ is the amplitude of transmission as hole-like quasiparticle in band 1, and $t^{he}_{12n}$ is the amplitude of transmission as hole-like quasiparticle from band 2. The coherence factors are given by, $u_{1(2)}= \sqrt[]{(1/2)(1+\Lambda_{1(2)}/E)}$, $v_{1(2)}= \sqrt[]{(1/2)(1-\Lambda_{1(2)}/E)} $, with $\Lambda_{1(2)} = \sqrt[]{E^2-\Delta_{1(2)}^2}$, $\Delta_1=|\Delta_1|$ and $\Delta_2=|\Delta_2|e^{-i\phi}$, $\phi$ represents the relative phase difference between band 1 and band 2. The wave function for the electron incident in band 2 is given by $e^{ik_1x}\phi^{N}_3$ instead of $e^{ik_1x}\phi^{N}_1$.
We have considered Andreev approximation such that Fermi energy $E_F\gg E,\Delta_1,\Delta_2$.
The wave vector of electron or hole in the normal metal is $k_{1(2)}=$  $k_F$, the wave vector of electron or hole like quasiparticle in Iron Pnictide is $q_{1e}=q_{1h}=q_{2e}=q_{2h}= k_F$, $k_F = \frac{\sqrt[]{2 m^* E_F}}{\hbar}$, $m^*$ is the effective mass of the electron and $E_F$ is the Fermi energy of the system\cite{PhysRevB.80.144507}. Throughout this work, we adopt the Andreev approximation. Nevertheless, we have verified that relaxing this approximation does not modify the qualitative features of the noise characteristics. For simplicity, the Fermi energy in both the normal metal and the Iron Pnictide superconductor is taken to be same, i.e., $E_F=10^4|\Delta_1|$, $|\Delta_1|=2.5meV$, and $|\Delta_2|=3.75meV$. The critical temperature $T_c$ for Iron Pnictide superconductor is around $18K$, see Ref.~\cite{Borisenko2010LiFeAs}.

Boundary conditions for setup shown in Fig~\ref{setup1}, with wavefunction defined in Eq.~\eqref{wavefunction} is given by:
% \hspace*{0.2cm}
% \begin{align}
% \Psi_{N}\big|_{x = 0} &= \Psi_{N_2}\big|_{x = -a}, \\[6pt]
% \left. \frac{\partial}{\partial x} \left( \Psi_{N_2} - \Psi_{N_1} \right) \right|_{x = -a} &=
% 2m^* \, V_1 \, \mathrm{diag}(\hat{1}, \hat{1}) \, \Psi_{N_1}\big|_{x = -a},
% \end{align}
\begin{subequations}\label{eq:bc}
\begin{align}
\Psi_{N}\big|_{x = 0} &= \Psi_{IP}\big|_{x = 0}, \text{and}
&& \label{eq:bc-a} \\[6pt]
\left. \frac{\partial}{\partial x}
\left( \Psi_{IP} - \Psi_{N} \right) \right|_{x = 0}
&=
\frac{2m^*}{\hbar^2} \left( V \, \mathrm{diag}(\hat{1}, \hat{1}) \right. \notag \\
&\quad \left. + {\tilde{\alpha}}_0 \, \mathrm{offdiag}(\hat{1}, \hat{1}) \right)
\Psi_{N}\big|_{x = 0}.
&& \label{eq:bc-b}
\end{align}
\end{subequations}

We take the dimensionless barrier strength to be $Z_{}=\frac{m^*V_{}}{\hbar^2 k_F}$ and the dimensionless inter-band coupling strength as ${\alpha}=\frac{m^*{\tilde{\alpha}}_{0}}{\hbar^2 k_F}$. $\hat{1}$ in Eq.~\eqref{eq:bc-b} represents the $2\times2$ unit matrix and diag($\hat{1}$,$\hat{1}$) and offdiag ($\hat{1}$,$\hat{1}$) represents the diagonal and off diagonal part of $4\times4$ block matrix respectively~\cite{PhysRevB.80.144507}. All scattering amplitudes required to construct the scattering matrix are obtained by considering the possible incidence of quasiparticle electrons and holes from both band~1 and band~2 onto the junction. A detailed calculation of the $S$-matrix has been given in Appendix ~\ref{s matrix}. Scattering states are constructed for electron and hole incidence from both metallic and superconducting sides, generating reflection and transmission amplitudes which form the elements of the scattering matrix $S$. The $S$-matrix is unitary ($S^\dagger S=I$), encoding normal reflection, Andreev reflection, and interband scattering processes. Probabilities for different channels are given by the squared moduli of the corresponding scattering amplitudes. A detailed calculation of the scattering matrix is given in Appendix \ref{s matrix}.

\par
We use the scattering matrix approach within the Blonder-Tinkham-Klapwijk (BTK) formalism to analyse charge transport, quantum noise properties in a normal-metal--insulator--Iron Pnictide (N-I-IP) junction. 

% The interface is characterised by a barrier strength dimensionless parameter $Z$ and interband coupling dimensionless parameter $\alpha$ as defined in Appendix \ref{s matrix}.

%@@@@@@@@@@@@@@@@@@@@@@@@@@@@@@@@@@@
\subsection{Current and differential conductance}
%@@@@@@@@@@@@@@@@@@@@@@@@@@@@@@@@@@@

The general expression for charge current is given by an energy integral over transmission probabilities weighted by Fermi functions  ~\cite{BENENTI20171}:
\begin{align}
I_{i,\gamma} &=
\frac{2e}{h} \sum_{j,\alpha}
\int_{0}^{\infty} dE
\operatorname{sgn}(\gamma)\big(
 N_i^\beta \delta_{i j} \delta_{\beta \alpha} - T_{i j}^{\beta \alpha} \big) f_{j \alpha}, 
\label{Ie}
\end{align}
where, $N_i^{\beta} = \sum\limits_{\substack{j, \gamma \in \{e,h\}}}T_{ij}^{\beta \gamma}$ and $T_{ij}^{\alpha \beta} = \sum\limits_{\substack{n,m \in {1,2}}}|s_{ij}^{\alpha\beta n m}|^2$, $s_{ij}^{\alpha\beta n m}$ are the scattering amplitudes for a particle of type $\beta$ scattered from band $m$ in terminal $j$ to terminal $i$ as a perticle $\alpha$ in band $n$. $f_{j\alpha}=$ $[1+e^{\frac{E+\operatorname{sgn}(\alpha)V_j}{k_BT_j}}]^{-1}$, is the Fermi function in contact $j$ for particle $\alpha$, $V_j$ is the applied voltage at terminal $j$, $k_B$ is the Boltzmann constant, and $T_j$ is temperature in terminal $j$.

In our setup, Fig.~\ref{setup1}, a voltage bias $V$ is applied to the normal metal, while the Iron Pnictide superconductor is kept grounded. The temperature of the normal metal is taken as $T_1=T+\frac{\Delta T}{2}$ and the temperature of the Iron Pnictide superconductor is $T_2=T-\frac{\Delta T}{2}$, such that the temperature bias across the junction is $\Delta T$. However, for the derivation of differential conductance, we keep $\Delta T = 0$. The current flowing through the normal metal terminal is given by
% From the sum rule, we will find out that $N_1^e=2$ \cite{Datta_1995}.
\begin{subequations}
\begin{align}
I_{N} &= I_{N,e}+I_{N,h}, \nonumber \\ 
&=
\frac{2e}{h} 
\int_{0}^{\infty} dE
\big[(2+T_{11}^{he}-T_{11}^{ee})(f_{1e}-f_{2e})\nonumber \\[6pt]
&\quad + (2+T_{11}^{eh}-T_{11}^{hh})(f_{2e}-f_{1h}) \big], 
\label{I1}
\end{align}
where
\begin{align}
\begin{split}
I_{N,e} &= \frac{2e}{h} 
\int_{0}^{\infty} dE
\big[(2+T_{11}^{he}-T_{11}^{ee})(f_{1e}-f_{2e})\big],\\
I_{N,h} &= \frac{2e}{h} 
\int_{0}^{\infty} dE
\big[(2+T_{11}^{eh}-T_{11}^{hh})(f_{2e}-f_{1h})\big].
\end{split}
\label{I10}
\end{align}
\end{subequations}

The corresponding Fermi-Dirac distribution functions are given by
\begin{align}
f_{1e} &= \frac{1}{e^{(E-eV)/k_B T}+1},\quad
f_{1h} = \frac{1}{e^{(E+eV)/k_B T}+1},
\nonumber\\
f_{2e} &= \frac{1}{e^{E/k_B T}+1}.
\end{align}

To obtain the differential conductance, we differentiate the total current $I_N$ with respect to the applied voltage bias $V$, i.e., $
dG_N = \frac{dI_N}{dV}.$ Since the total current consists of electron and hole contributions, the differential conductance can also be decomposed into electron and hole parts as,
\begin{align}
\frac{dI_N}{dV}
=
\frac{dI_{N,e}}{dV}
+
\frac{dI_{N,h}}{dV} = dG_{e} + dG_{h}.
\end{align}

Differentiating Eq.~\eqref{I10} with respect to the voltage bias $V$, we obtain
\begin{subequations}
\begin{align}
dG_e
&=
\frac{dI_{N,e}}{dV}
\nonumber\\[4pt]
&=
\frac{2e}{h}
\int_{0}^{\infty} dE
\,
(2+T_{11}^{he}-T_{11}^{ee})
\big(f'_{1e}-f'_{2e}\big),
\\[8pt]
dG_h
&=
\frac{dI_{N,h}}{dV}
\nonumber\\[4pt]
&=
\frac{2e}{h}
\int_{0}^{\infty} dE
\,
(2+T_{11}^{eh}-T_{11}^{hh})
\big(f'_{2e}-f'_{1h}\big),
\end{align}
\label{eq:dG_intermediate}
\end{subequations}
where $f'_{jq}=\frac{\partial f_{jq}}{\partial V}$ for $j \in \{1,2 \}$ and $q \in \{e,h \}$.
We note that only $f_{1e}$ and $f_{1h}$ depend explicitly on the voltage bias $V$, whereas $f_{2e}$ is completely independent of $V$. Therefore,
\begin{align}
\begin{split}
\big(f'_{1e}-f'_{2e}\big)
&=
{f'_{1e}},
\quad
\big(f'_{2e}-f'_{1h}\big)
=
-f'_{1h}.
\end{split}
\end{align}

Using the identities of the Fermi-Dirac distribution functions,
\begin{align}
\begin{split}
f'_{1e}
&=
-e\frac{\partial f_{1e}}{\partial E},
\quad
f'_{1h}
=
e\frac{\partial f_{1h}}{\partial E},
\end{split}
\end{align}
the finite-temperature differential conductance can then be written as
\begin{subequations}
\begin{align}
dG_e
&=
\frac{2e^2}{h}
\int_{0}^{\infty} dE
\,
(2+T_{11}^{he}-T_{11}^{ee})
\left(
-\frac{\partial f_{1e}}{\partial E}
\right),
\label{eq:finite_temp_dG1}
\\[8pt]
dG_h
&=
\frac{2e^2}{h}
\int_{0}^{\infty} dE
\,
(2+T_{11}^{eh}-T_{11}^{hh})
\left(
-\frac{\partial f_{1h}}{\partial E}
\right).
\label{eq:finite_temp_dG}
\end{align}
\end{subequations}

Thus, total differential conductance at any arbitrary temperature is given by $dG_N=dG_e+dG_h.$ At zero temperature, the derivative of the Fermi-Dirac distribution reduces to a Dirac delta function. Therefore,
\begin{align}
-\frac{\partial f_{1e}}{\partial E}
=
\delta(E-eV),
\qquad
-\frac{\partial f_{1h}}{\partial E}
=
\delta(E+eV).
\label{eq:delta_identity}
\end{align}

For positive bias, i.e., $eV>0$, the delta function $\delta(E-eV)$ lies within the integration window $0<E<eV$. Consequently, only the electron contribution survives in this regime. Thus, Eq.~\eqref{eq:finite_temp_dG1} reduces to
\begin{align}
dG_e
&=
\frac{2e^2}{h}
\int_{0}^{eV} dE
\,
(2+T_{11}^{he}-T_{11}^{ee})
\delta(E-eV).
\end{align}

Using the property of the Dirac delta function for an arbitrary function $F(E)$,
\begin{align}
\int dE\,F(E)\delta(E-eV)=F(eV),
\label{eq:14}
\end{align}
the above expression simplifies to,
\begin{align}
dG_e
&=
\frac{2e^2}{h}
\left[
2+T_{11}^{he}(eV)-T_{11}^{ee}(eV)
\right],
\qquad \text{for } eV>0.
\label{eq:dGe_zeroT}
\end{align}

Similarly, at zero-temperature, $dG_h$ as in Eq.~\eqref{eq:finite_temp_dG} reduces to,
\begin{align}
dG_h
&=
\frac{2e^2}{h}
\int_{0}^{\infty} dE
\,
(2+T_{11}^{eh}-T_{11}^{hh})
\delta(E+eV).
\label{eq:Gh1}
\end{align}

On the other hand, for positive bias $eV>0$, the delta function $\delta(E+eV)$ is centered at $E=-eV$, which lies outside the integration range $0<E<\infty$. Therefore, the hole contribution vanishes for positive bias. However, for negative bias, i.e., $eV<0$, the center of the delta function shifts to $E=-eV>0$, which now lies within the allowed integration window. Consequently, the hole contribution to the differential conductance becomes finite. To express the hole contribution in terms of negative energies, we perform the transformation $
E \rightarrow -E.$
Under this transformation, the integration limits change from ($0$, $\infty$) for $E$ to ($0$, $-\infty$) for $-E$.
Reversing the order of integration introduces an additional minus sign, which cancels with the minus sign arising from the differential transformation $dE \rightarrow -dE$. Consequently, Eq.~\eqref{eq:Gh1} reduces to,
\begin{align}
dG_h
&=
\frac{2e^2}{h}
\int_{-\infty}^{0} dE
\,
(2+T_{11}^{eh}-T_{11}^{hh})
\delta(-E+eV).
\end{align}

Using the property of the Dirac delta function, $\delta(-x)=\delta(x),$, we obtain $\delta(-E+eV)
=
\delta(E-eV).$

Therefore, the hole contribution can be written as,
\begin{align}
dG_h
&=
\frac{2e^2}{h}
\int_{-\infty}^{0} dE
\,
(2+T_{11}^{eh}-T_{11}^{hh})
\delta(E-eV).
\label{eq:Gh2}
\end{align}

Since $eV<0$, the delta function $\delta(E-eV)$ is centered at a negative energy $eV<0$ lying within the integration window $-\infty<E<0$. Hence, the integral gives a finite contribution. Restricting the integration window around the location of the delta peak, Eq.~\eqref{eq:Gh2} can equivalently be written as
\begin{align}
dG_h
&=
\frac{2e^2}{h}
\int_{eV}^{0} dE
\,
(2+T_{11}^{eh}-T_{11}^{hh})
\delta(E-eV).
\end{align}

Finally, using the delta-function identity as in Eq. (\ref{eq:14}), the hole contribution to the differential conductance at zero temperature becomes
\begin{align}
dG_h
&=
\frac{2e^2}{h}
\left[
2+T_{11}^{eh}(eV)-T_{11}^{hh}(eV)
\right],
\qquad \text{for } eV<0.
\label{eq:dGh_zeroT}
\end{align}

Therefore, the zero-temperature differential conductance is given by,

\begin{align}
\begin{split}
dG_e
&=
\frac{2e^2}{h}
\left[
2+T_{11}^{he}(eV)-T_{11}^{ee}(eV)
\right],
\qquad \text{for } eV>0,
\\
dG_h
&=
\frac{2e^2}{h}
\left[
2+T_{11}^{eh}(eV)-T_{11}^{hh}(eV)
\right],
\qquad \text{for } eV<0.
\end{split}
\label{eq:zero_temp_dG_final}
\end{align}

Here, the differential conductance ($dG_N$) has been derived in its most general form without imposing any restriction on the applied voltage bias or temperature gradient. The obtained expression remains valid both in the low-bias and high-bias regimes. However, the analytical expression for the conductance ($G$) can be derived in a much simpler form within the linear response regime, where the applied voltage bias and temperature difference satisfy $eV,\, \Delta T \ll k_B T$. In this limit, the Fermi-Dirac distribution functions can be expanded around equilibrium, which considerably simplifies the transport equations. Beyond the linear response regime, higher-order nonequilibrium contributions become important, making the analytical derivation of conductance significantly more complicated.

% For finite $Z$ and $\alpha$, electron-hole asymmetry appears due to interband coupling, see Fig. \ref{con22}.

% \par
% The current in the normal metal–insulator–normal metal (N–I–N) junction is thus,
% \begin{equation}
%     I_{0} = \frac{2e}{h} \int_{0}^{\infty} dE \,
%     \Big[ (2 - T_{11,n}^{ee})(f_{1e} - f_{2e})
%     + (2 - T_{11,n}^{hh})(f_{1h} - f_{2e}) \Big],
% \end{equation}
% where $T_{11,n}^{ee}$ and $T_{11,n}^{hh}$ denote the probabilities of normal reflection in a N-I-N junction. 
% The excess current is defined as,
% \begin{equation}
%     I_{exc} = I_{N} - I_{0},
% \end{equation}
% and at zero temperature ($T=0$) for $eV>0$, it reduces to
% \begin{equation}
%     I_{exc} = \frac{1+Z^2}{2eR_N}\,\int_0^{eV}dE\big[T_{11}^{he} - T_{11}^{ee} + T_{11,n}^{ee}\big]\,,
% \end{equation}

%@@@@@@@@@@@@@@@@@@@@@@@@@@@@@@@@
\subsection{Conductance}
\label{sec2c}
%@@@@@@@@@@@@@@@@@@@@@@@@@@@@@@@@

\subsubsection{Current in Linear response regime}
Here, we derive the conductance ($G$) and Seeback coefficient ($S$) considering only the linear response regime, i.e., ($eV,\, \Delta T \ll T$), we start from the equation of current as in Eq. (\ref{I1}), i.e.,

\begin{align}
I_{N} &= I_{N,e}+I_{N,h}, \nonumber \\ 
&=
\frac{2e}{h} 
\int_{0}^{\infty} dE
\big[(2+T_{11}^{he}-T_{11}^{ee})(f_{1e}-f_{2e})\nonumber \\[6pt]
&\quad + (2+T_{11}^{eh}-T_{11}^{hh})(f_{2e}-f_{1h}) \big], 
\label{eq:21}
\end{align}

In the linear-response regime 
($eV,\, \Delta T \ll T$), one can expand the Fermi function to linear order in $eV$ and $\Delta T$, i.e.,
\begin{equation}
\begin{split}
f_{1e}
= f_0
- eV \frac{\partial f_0}{\partial E}
- \frac{\Delta T}{2}\frac{E}{T}
\frac{\partial f_0}{\partial E}, \\ f_{1h}
= f_0
+ eV \frac{\partial f_0}{\partial E}
- \frac{\Delta T}{2}\frac{E}{T}
\frac{\partial f_0}{\partial E},\\ f_{2e}(E,T_2)
= f_0
+ \frac{\Delta T}{2}\frac{E}{T}
\frac{\partial f_0}{\partial E}.
\end{split}
\label{Taylor}
\end{equation}

Therefore, $f_{1e} - f_{2e}
= - eV \frac{\partial f_0}{\partial E}
- \frac{\Delta T}{T} E \frac{\partial f_0}{\partial E}, \quad f_{2e} - f_{1h}
= \frac{\Delta T}{T} E \frac{\partial f_0}{\partial E}
- eV \frac{\partial f}{\partial E},$ where $f_0 = \frac{1}{1+\exp\!\left(\frac{E}{k_B T}\right)}.$

Substituting into Eq.~\eqref{eq:21} and collecting terms,
\begin{align}
\begin{split}
I_N
&= \frac{2e}{h} \int_0^\infty dE \,
\Big[
(4 + T_{11}^{he} + T_{11}^{eh} - T_{11}^{hh} - T_{11}^{ee})
eV \left(-\frac{\partial f_0}{\partial E}\right) \\&+
(T_{11}^{he} - T_{11}^{ee} - T_{11}^{eh} + T_{11}^{hh})
\frac{\Delta T}{T} E
\left(-\frac{\partial f_0}{\partial E}\right)
\Big],
\end{split}
\label{I1linear}
\end{align}

 Eq.~\eqref{I1linear} can be written in the standard linear-response form
\begin{equation}
I_N = G V + L \Delta T ,
\label{eq:28}
\end{equation}
where,
\begin{subequations} \label{eq:I_cases}
\begin{align}
    G &= \frac{2e^2}{h}
\int_{0}^{\infty} dE
(4+T_{11}^{eh}+T_{11}^{he}-T_{11}^{hh}-T_{11}^{ee})\left(-\frac{\partial f_0}{\partial E}\right),\label{gfinite}  \\
    L &= \frac{2e}{h}
\int_{0}^{\infty} dE
(T_{11}^{he}-T_{11}^{ee}-T_{11}^{eh}+T_{11}^{hh})\frac{E}{T}\left(-\frac{\partial f_0}{\partial E}\right),
\label{eq:I_neg}
\end{align}
\end{subequations}
with $G$ being the conductance in linear response and $L$ being the Onsager coefficient. The Seeback coefficient ($S$) is defined as $\frac{L}{G}$. {However, since both the normal metal and iron-pnictide superconductor are electron-hole symmetric, $T_{11}^{he} = T_{11}^{eh}$ and $T_{11}^{hh} = T_{11}^{ee}$, which is why $L$ is zero, which means the Seeback coefficient $S$ is also zero. Therefore, the zero current, i.e., $I_N = 0$ is achieved at zero thermovoltage.}

%@@@@@@@@@@@@@@@@@@@@@@@@@@@@@@@
\subsection{Finite temperature quantum noise}
\label{quantum noise}
%@@@@@@@@@@@@@@@@@@@@@@@@@@@@@@@

The current-current correlation between different terminals is defined as quantum noise. Quantum noise correlation \cite{Blanter2000} between terminals $i$ and $j$ with particle types $\alpha$ and $\beta$ at times $t$ and $t'$ is,
\begin{equation}
Q^{}_{\alpha \beta,ij}(t - t') = \langle \Delta I^{\alpha}_i(t) \Delta I^{\beta}_j(t') + \Delta I^{\beta}_j(t') \Delta I^{\alpha}_i(t) \rangle,
\label{timefluctuaions1}
\end{equation}
where $\Delta I^{\alpha}_i(t) = I^{\alpha}_i(t) - \langle I^{\alpha}_i(t) \rangle$ denotes the fluctuation in current of particle $\alpha$ ( $\alpha \in \{e,  h\}$) with $i,j\in \{N\}$ denoting terminals \cite{datta1996} as we are interested in the autocorrelation. The Fourier transform of Eq. \eqref{timefluctuaions1} gives the frequency-dependent current correlations, and one can calculate the correlation between particles at zero frequency, i.e., 
\begin{align}
Q_{\alpha \beta, ij}(\omega = 0) 
&= \sum_{\substack{m,n \in \{1,2\}}} Q^{mn}_{\alpha \beta, ij}(\omega=0),
\label{eqn:Qnoise1}
\end{align}
\begin{widetext}
with
\begin{align}
{Q^{mn}_{\alpha \beta,i j} (\omega =0)} &=  \frac{e^2}{h} 
\int \sum_{\substack{k,l \in \{1, 2\}, \\ \gamma,\delta \in \{e,h\}\\ m',n' \in \{1,2\}}}
sgn(\alpha) sgn(\beta) A_{k,\gamma;l,\delta}^{m'n'}(i,  \alpha,E) A_{l,\delta ;k,\gamma}^{n'm'}(j , \beta, E)   \times [ \textit{f}_{k \gamma}(E) (1-\textit{f}_{l \delta}(E)) +  \textit{f}_{l \delta}(E) (1-\textit{f}_{k \gamma}(E))] dE, 
\label{eqn:Qnoise}
\end{align}
\end{widetext}

wherein $A_{k,\gamma;l,\delta}^{m'n'}(i,  \alpha,E)=\delta_{ik}\delta_{il}\delta_{\alpha\gamma}\delta_{\alpha \delta}\delta_{m' n'}-s_{ik}^{\alpha\gamma mm'^\dagger} s_{il}^{\alpha\delta mn'}$.

The quantum noise autocorrelation is then given by,

\begin{equation}
Q_{11} = Q_{ee,11}+ Q_{eh,11}+ Q_{he,11}+ Q_{hh,11}.
\label{qnoisemain}
\end{equation}

The finite temperature quantum {equilibrium noise} is thus derived as,
% \begin{align}
% Q_{11}^{sh} &= \frac{2e^2}{h} \int_0^\infty dE\; \left[ c_1 (f_{1e}-f_{1h})^2 + c_2 (f_{1e}-f_{2e})^2 + c_3 (f_{1h}-f_{2e})^2 \right]
% \label{qsh11}
% \end{align}
\begin{align}
Q_{11}^{\mathrm{eq}}
&= \frac{2e^{2}}{h}
\int_{0}^{\infty} dE\;
\Big[
t_{1}\,f_{1h}(1-f_{1h})
+ t_{2}\,f_{1e}(1-f_{1e})
\\ \nonumber
&\hspace{3.5em}
+ t_{3}\,f_{2e}(1-f_{2e})
\Big],
\label{qshm1}
\end{align}

where $t_1,t_2$ and $t_3$ are derived in Appendix ~\ref{quantum noise appendix}. First, using Eq.~(\ref{Taylor}), we define the deviations from equilibrium distribution \(f_0\) as,
\begin{align}
f_{1e}=f_0+\delta f_{1e},
\qquad
f_{1h}=f_0+\delta f_{1h},
\qquad
f_{2e}=f_0+\delta f_{2e},
\end{align}
where,
\begin{align}
\begin{split}
\delta f_{1e}
=
- eV \frac{\partial f_0}{\partial E}
-\frac{\Delta T}{2}\frac{E}{T}\frac{\partial f_0}{\partial E},\\
\delta f_{1h}
=
+ eV \frac{\partial f_0}{\partial E}
-\frac{\Delta T}{2}\frac{E}{T}\frac{\partial f_0}{\partial E},\\
\delta f_{2e}
=
+\frac{\Delta T}{2}\frac{E}{T}\frac{\partial f_0}{\partial E}.
\end{split}
\label{eq:36}
\end{align}

Neglecting second and higher-order terms in \(\delta f_{1e}\), \(\delta f_{1h}\) and \(\delta f_{2e}\), we obtain,
\begin{align}
\begin{split}
f_{1e}(1-f_{1e})
=
f_0(1-f_0)
+(1-2f_0)\delta f_{1e},\\
f_{1h}(1-f_{1h})
=
f_0(1-f_0)
+(1-2f_0)\delta f_{1h},\\
f_{2e}(1-f_{2e})
=
f_0(1-f_0)
+(1-2f_0)\delta f_{2e}.
\end{split}
\end{align}

Substituting the explicit forms of \(\delta f_{1e}\), \(\delta f_{1h}\), and \(\delta f_{2e}\) from Eq. (\ref{eq:36}), we obtain,
\begin{align}
\begin{split}
f_{1e}(1-f_{1e})
&=
f_0(1-f_0)
+(1-2f_0)
\left[
- eV \frac{\partial f_0}{\partial E}
-\frac{\Delta T}{2}\frac{E}{T}
\frac{\partial f_0}{\partial E}
\right],
\\
f_{1h}(1-f_{1h})
&=
f_0(1-f_0)
+(1-2f_0)
\left[
+ eV \frac{\partial f_0}{\partial E}
-\frac{\Delta T}{2}\frac{E}{T}
\frac{\partial f_0}{\partial E}
\right],
\\
f_{2e}(1-f_{2e})
&=
f_0(1-f_0)
+(1-2f_0)
\left[
+\frac{\Delta T}{2}\frac{E}{T}
\frac{\partial f_0}{\partial E}
\right].
\end{split}
\label{eq:38}
\end{align}

Substituting Eq. (\ref{eq:38}) in Eq.~(\ref{qshm1}), the {equilibrium noise} becomes
\begin{align}
\begin{split}
Q_{11}^{\mathrm{eq}}
&=
\frac{2e^2}{h}
\int_0^\infty dE
\Bigg\{
t_1
\Bigg[
f_0(1-f_0)
+(1-2f_0)\\&
\times \left(
eV\frac{\partial f_0}{\partial E}
-
\frac{\Delta T}{2}\frac{E}{T}
\frac{\partial f_0}{\partial E}
\right)
\Bigg]
+t_2
\Bigg[
f_0(1-f_0)\\&
+(1-2f_0)
\left(
- eV\frac{\partial f_0}{\partial E}
-
\frac{\Delta T}{2}\frac{E}{T}
\frac{\partial f_0}{\partial E}
\right)
\Bigg]
\\
&
+t_3
\Bigg[
f_0(1-f_0)
+(1-2f_0)
\left(
\frac{\Delta T}{2}\frac{E}{T}
\frac{\partial f_0}{\partial E}
\right)
\Bigg]
\Bigg\}.
\end{split}
\end{align}

Simplifying further, we finally obtain,
\begin{align}
Q_{11}^{\mathrm{eq}}
&=
\frac{2e^2}{h}
\int_0^\infty dE
\Bigg[
(t_1+t_2+t_3)f_0(1-f_0)
\nonumber\\
&\qquad
+eV(t_1-t_2)
(1-2f_0)\frac{\partial f_0}{\partial E}
\nonumber\\
&\qquad
-\frac{\Delta T}{2}\frac{E}{T}
(t_1+t_2-t_3)
(1-2f_0)\frac{\partial f_0}{\partial E}
\Bigg].
\label{eq:Qth}
\end{align}

The general expression for quantum shot noise irrespective of whether one is at zero temperature or finite temperature is then,
% \begin{align}
% Q_{11}^{sh} &= \frac{2e^2}{h} \int_0^\infty dE\; \left[ c_1 (f_{1e}-f_{1h})^2 + c_2 (f_{1e}-f_{2e})^2 + c_3 (f_{1h}-f_{2e})^2 \right]
% \label{qsh11}
% \end{align}
\begin{equation}
\begin{split}
Q_{11}^{\mathrm{sh}}
&= \frac{2e^{2}}{h}
\int_{0}^{\infty} dE\;
\Big[
c_{1}\,(f_{1e}-f_{1h})^{2}
+ c_{2}\,(f_{1e}-f_{2e})^{2}
\\
&\hspace{3.5em}
+ c_{3}\,(f_{1h}-f_{2e})^{2}
\Big],
\end{split}
\label{qshm}
\end{equation}

where $c_1,c_2$ and $c_3$ are derived in Appendix ~\ref{quantum noise appendix}. Starting from Eq.~(\ref{qshm}), we now substitute the linear response expansion of the Fermi distribution functions obtained from Eq.~(\ref{Taylor}). First, the differences between the distribution functions are given by
\begin{equation}
\begin{split}
f_{1e}-f_{1h}
&=
-2eV\frac{\partial f_0}{\partial E},
\\
f_{1e}-f_{2e}
&=
-\left(
eV+\frac{\Delta T\,E}{T}
\right)
\frac{\partial f_0}{\partial E},
\\
f_{1h}-f_{2e}
&=
\left(
eV-\frac{\Delta T\,E}{T}
\right)
\frac{\partial f_0}{\partial E}.
\end{split}
\end{equation}

Therefore,
\begin{align}
\begin{split}
(f_{1e}-f_{1h})^2
=
4e^2V^2
\left(
\frac{\partial f_0}{\partial E}
\right)^2,\\
(f_{1e}-f_{2e})^2
=
\left(
eV+\frac{\Delta T\,E}{T}
\right)^2
\left(
\frac{\partial f_0}{\partial E}
\right)^2,\\
(f_{1h}-f_{2e})^2
=
\left(
eV-\frac{\Delta T\,E}{T}
\right)^2
\left(
\frac{\partial f_0}{\partial E}
\right)^2.
\end{split}
\end{align}

Substituting these expressions into Eq.~(\ref{qshm}), the finite-temperature quantum shot noise becomes
\begin{align}
\begin{split}
Q_{11}^{\mathrm{sh}}
&=
\frac{2e^2}{h}
\int_0^\infty dE \left(
\frac{\partial f_0}{\partial E}
\right)^2
\Bigg[
4c_1e^2V^2
+c_2
\left(
eV+\frac{\Delta T\,E}{T}
\right)^2
\\
&\qquad
+c_3
\left(
eV-\frac{\Delta T\,E}{T}
\right)^2
\Bigg].
\end{split}
\end{align}

Further simplifying, we obtain
\begin{align}
\begin{split}
Q_{11}^{\mathrm{sh}}
&=
\frac{2e^2}{h}
\int_0^\infty dE
\left(
\frac{\partial f_0}{\partial E}
\right)^2
\Bigg[
4c_1e^2V^2
\\
&\qquad
+c_2
\left(
e^2V^2
+
2eV\frac{\Delta T\,E}{T}
+
\frac{\Delta T^2E^2}{T^2}
\right)
\\
&\qquad
+c_3
\left(
e^2V^2
-
2eV\frac{\Delta T\,E}{T}
+
\frac{\Delta T^2E^2}{T^2}
\right)
\Bigg].
\end{split}
\end{align}

Finally, collecting similar terms, the finite-temperature quantum shot noise reduces to
\begin{align}
Q_{11}^{\mathrm{sh}}
&=
\frac{2e^2}{h}
\int_0^\infty dE
\left(
\frac{\partial f_0}{\partial E}
\right)^2
\Bigg[
e^2V^2(4c_1+c_2+c_3)
\nonumber\\
&\qquad
+2eV\frac{\Delta T\,E}{T}(c_2-c_3)
+\frac{\Delta T^2E^2}{T^2}(c_2+c_3)
\Bigg].
\label{Qsh}
\end{align}

% Only the squared differences of Fermi functions contribute to the shot noise component, while other terms yield {equilibrium noise}, see Appendix~\ref {quantum noise appendix} for the full derivation of Eq.~\ref{qshm}.

Finite temperature quantum noise is the addition of finite temperature quantum shot noise and finite temperature {equilibrium noise}, i.e., $Q_{11}=Q_{11}^{\mathrm{eq}}+Q_{11}^{\mathrm{sh}}.$ In our work, for the calculation of finite-temperature quantum noise $Q_{11}$, we consider $eV \neq 0, \Delta T = 0$, therefore, the final expressions of $Q_{11}^{\text{th}}$ and $Q_{11}^{\text{sh}}$ are given as

\begin{equation}
    \begin{split}
       Q_{11}^{\mathrm{eq}}
&=
\frac{2e^2}{h}
\int_0^\infty dE
\Bigg[
(t_1+t_2+t_3)f_0(1-f_0)
\\
&\qquad
+eV(t_1-t_2)
(1-2f_0)\frac{\partial f_0}{\partial E},\\
Q_{11}^{\mathrm{sh}}
&=
\frac{2e^2}{h}
\int_0^\infty dE
\left(
\frac{\partial f_0}{\partial E}
\right)^2
\Bigg[
e^2V^2(4c_1+c_2+c_3)\Bigg].
    \end{split}
    \label{eq:45}
\end{equation}

%@@@@@@@@@@@@@@@@@@@@@@@@@@@@@@@@
\subsubsection{Zero temperature quantum shot noise}
%@@@@@@@@@@@@@@@@@@@@@@@@@@@@@@@@

% At $T=0$ and $V>0$, the form of Fermi functions in the quantum shot formula, which contribute to quantum shot noise are $(f_{1h}-f_{1e})^2$ and $(f_{1e}-f_{2e})^2$. At $T=0K$, both of these terms remain unity up to the applied voltage bias $V$. Hence, the formula for shot noise becomes,

At the zero-temperature limit and finite-bias regime, i.e., $T_1=T_2=0$ K and finite $V$, the total quantum noise is entirely governed by the quantum shot-noise contribution. In this limit, the Fermi-Dirac distribution functions reduce to Heaviside step functions. Consequently, for positive bias ($eV>0$), the electron distribution function in the normal-metal terminal becomes
\begin{align}
f_{1e}(E)=
\begin{cases}
1, & 0<E<eV,\\
0, & E>eV.
\end{cases}
\end{align}

Similarly, the equilibrium distribution function of the grounded superconducting terminal is given by
\begin{align}
f_{2e}(E)=
\begin{cases}
1, & E<0,\\
0, & E>0.
\end{cases}
\end{align}

For negative bias ($eV<0$), the hole distribution function satisfies
\begin{align}
f_{1h}(E)=
\begin{cases}
1, & eV<E<0,\\
0, & \text{otherwise}.
\end{cases}
\end{align}. Therefore, the differences of the Fermi-Dirac distribution functions appearing in Eq.~(\ref{qshm}) simplify considerably in the zero-temperature limit. In particular, for positive bias ($eV>0$),
\begin{align}
(f_{1e}-f_{2e})=
\begin{cases}
1, & 0<E<eV,\\
0, & \text{otherwise},
\end{cases}
\end{align}
since $f_{1e}=1$ and $f_{2e}=0$ within the transport window $0<E<eV$.

Similarly, for negative bias ($eV<0$),
\begin{align}
(f_{2e}-f_{1h})=
\begin{cases}
-1, & eV<E<0,\\
0, & \text{otherwise},
\end{cases}
\end{align}
while
\begin{align}
(f_{1e}-f_{1h})=
\begin{cases}
1, & 0<E<eV,\\
0, & \text{otherwise}.
\end{cases}
\end{align}

Since we consider only positive bias throughout the manuscript ($eV>0$), the hole contribution vanishes identically within the transport window. Consequently, the term proportional to $c_3$, i.e., $(f_{1h}-f_{2e})^2$, does not contribute to the quantum shot noise. On the other hand, within the interval $0<E<eV$, the remaining distribution-function differences satisfy
\begin{align}
(f_{1e}-f_{1h})^2=
\begin{cases}
1, & 0<E<eV,\\
0, & \text{otherwise},
\end{cases}
\end{align}
and
\begin{align}
(f_{1e}-f_{2e})^2=
\begin{cases}
1, & 0<E<eV,\\
0, & \text{otherwise}.
\end{cases}
\end{align}

Therefore, only the coefficients $c_1$ and $c_2$ contribute within the transport window, and Eq.~(\ref{qshm}) reduces to
\begin{align}
\bar{Q}_{11}^{\mathrm{sh}}
=
\frac{2e^2}{h}
\int_0^{eV} dE
\left[
c_1+c_2
\right].
\end{align}

It is important to emphasize that the expanded expression of $Q_{11}^{\mathrm{sh}}$ in Eq.~(\ref{eq:45}) was derived within the linear-response regime under the condition $
eV \ll k_BT.$
However, quantum shot noise becomes the dominant contribution in the opposite limit,
$
eV \gg k_BT,$
which corresponds to the nonequilibrium finite-bias regime. The zero-temperature quantum shot noise $\bar{Q}_{11}^{\mathrm{sh}}$ therefore belongs to this large-bias limit, where the linear-response expansion of the Fermi-Dirac distribution functions is no longer valid. Consequently, Eq.~(\ref{eq:45}) cannot be directly used to describe the zero-temperature quantum shot noise.

\subsection{$\Delta_T$ noise}
In the linear response regime, the current is given by  Eq. ~\eqref{eq:28}. In the case of a N–I–IP junction, the setup is intrinsically electron–hole {symmetric}. Therefore, in an electron–hole symmetric system, the thermoelectric response vanishes, and consequently the thermovoltage is identically zero under open-circuit (zero-current) conditions. Under this condition, the measured quantum shot noise corresponds to the $\Delta_T$ noise, thus, 
% The detailed derivation of the corresponding expression is presented in 
% Appendix~\ref{dtnoise appendix}.

\begin{align}
\Delta_{T} &=
\frac{2e^2}{h} \int_0^\infty dE\;
\Big[
c_1 (f_{1e}-f_{1h})^2
+ c_2 (f_{1e}-f_{2e})^2 \notag \\
&\quad
+ c_3 (f_{1h}-f_{2e})^2
\Big].
\label{dtnoise}
\end{align}

In the linear response regime, $\Delta_T$ noise at $V = V_{th} = 0$ is given as,

\begin{align}
\Delta_{T}
&=
\frac{2e^2}{h}
\int_0^\infty dE
\left(
\frac{\partial f_0}{\partial E}
\right)^2
\Bigg[
\frac{\Delta T^2E^2}{T^2}(c_2+c_3)
\Bigg].
\label{Qsh}
\end{align}

%#############################################################
%#############################################################
%#############################################################
%#############################################################

\section{Results and Discussion}
\label{results}

% The conductance fails to distinguish the pairing symmetries properly; the results are presented in the Appendix. Therefore, we look at quantum shot noise and $\Delta_T$ noise.
We first discuss the conductance, followed by 
zero-temperature quantum shot noise and finite-temperature quantum noise. We then present $\Delta_T$ noise. All calculations presented in this work were performed using \textit{Mathematica}. The code used for the numerical simulations is publicly available at the GitHub repository \cite{githublink}.

\subsection{Differential Conductance and Conductance}

%___________________________-

The differential conductance ($dG_N$) for different interband coupling strengths $\alpha$ is shown in Fig.~\ref{conv} at zero temperature with barrier strength $Z=1$ utilizing Eq. (\ref{eq:zero_temp_dG_final}). Panels (a)–(d) correspond to $\alpha = 0, 1, 2,$ and $3$, respectively. For all the values of $\alpha$, the differential conductance spectra for both $S_{++}$ and $S_{+-}$ pairing symmetries are symmetric with respect to the applied voltage bias.

\begin{widetext}

    \begin{figure}[H]
        \centering
        \includegraphics[width=1.0\linewidth]{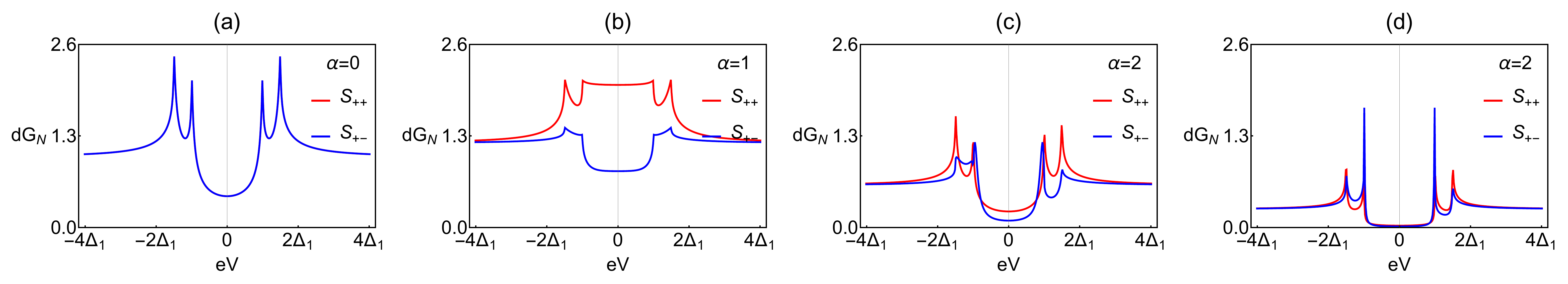}
        \caption{Differential Conductance $dG_N$ (in units of $\tfrac{2e^2}{h}$) as a function of $eV$ at $T_1=T_2=0K$, Z=1 for different interband coupling strengths($\alpha$). Panels (a), (b), (c) and (d) correspond 
to $\alpha = 0$, $1$, $2$ and $3$, respectively, showing the conductance spectra for the 
$S_{++}$ pairing symmetry (red) and the $S_{+-}$ pairing symmetry (blue).}
        \label{conv}
    \end{figure}
\end{widetext}

At finite temperatures $T_1 = 11\,\mathrm{K}$ and $T_2 = 9\,\mathrm{K}$, the 
conductance is given by Eq.~\eqref{gfinite}. In Fig.~\ref{conz}, we plot the 
conductance as a function of the barrier strength $Z$ for different values of 
the interband coupling strength $\alpha$.

For $\alpha = 0$, both pairing symmetries exhibit identical conductance 
behaviour, characterized by a pronounced peak in the transparent limit 
($Z \approx 0$), followed by a monotonic decay towards zero in the 
tunnelling regime. At $\alpha = 1$, a qualitative distinction begins to emerge: the 
$S_{++}$ pairing symmetry develops a pronounced peak around 
$Z \approx 1$, whereas the $S_{+-}$ state also shows a peak in the 
same region but with a noticeably smaller magnitude. When the interband coupling is increased further to $\alpha = 2$, both pairing symmetries display very low conductance in the transparent 
regime. As the barrier strength increases, the conductance develops a 
peak near $Z \approx \alpha$, followed by a decay in the tunnelling 
limit. For stronger interband coupling, $\alpha = 3$, a similar qualitative 
trend is observed; however, the conductance maximum shifts to higher 
barrier strength, appearing near $Z \approx 3$. Although both pairing symmetry shows a single peak around $Z\approx\alpha$, only magnitudes are different but experimentally which is not a good probe hence studying second moment noise is essntial.

\begin{widetext}

    \begin{figure}[H]
        \centering
        \includegraphics[width=1.0\linewidth]{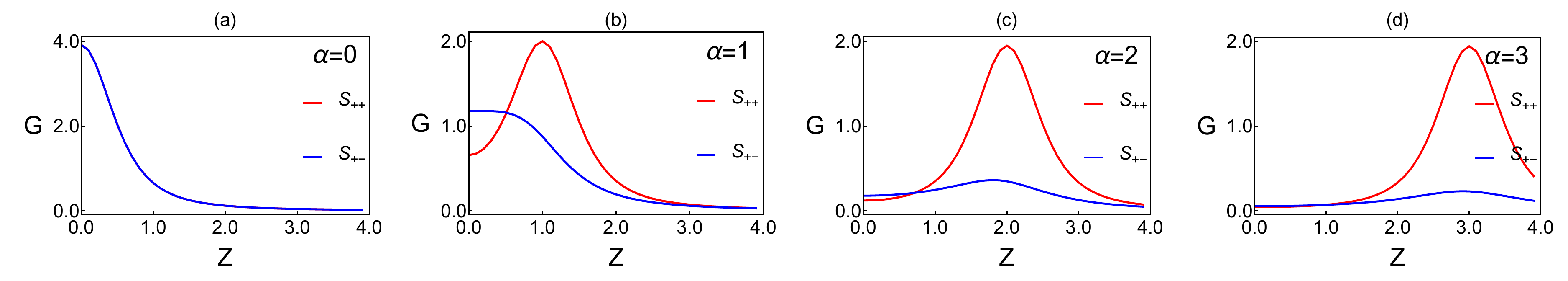}
        \caption{Conductance $G$ (in units of $\tfrac{2e^2}{h}$) as a function of $Z$ at $T=10K$ for different interband coupling strengths($\alpha$). Panels (a), (b), (c) and (d) correspond 
to $\alpha = 0$, $1$, $2$ and $3$, respectively, showing the conductance spectra for the 
$S_{++}$ pairing symmetry (red) and the $S_{+-}$ pairing symmetry (blue).}
        \label{conz}
    \end{figure}
\end{widetext}

In Fig.~\ref{cona}, we plot the conductance as a function of the 
interband coupling strength $\alpha$ for fixed barrier strengths 
$Z = 0$, $1$, $2$, and $3$. 

For $Z = 0$, both pairing symmetries exhibit identical behaviour, with 
a pronounced peak at $\alpha \approx 0$ followed by a monotonic decrease 
as $\alpha$ increases. For finite barrier strengths, the conductance is 
small at weak interband coupling and develops a peak near 
$\alpha \approx Z$, after which it decreases again. As $Z$ increases, 
the position of the conductance maximum shifts systematically to higher 
values of $\alpha$.

\begin{widetext}

    \begin{figure}[H]
        \centering
        \includegraphics[width=1.0\linewidth]{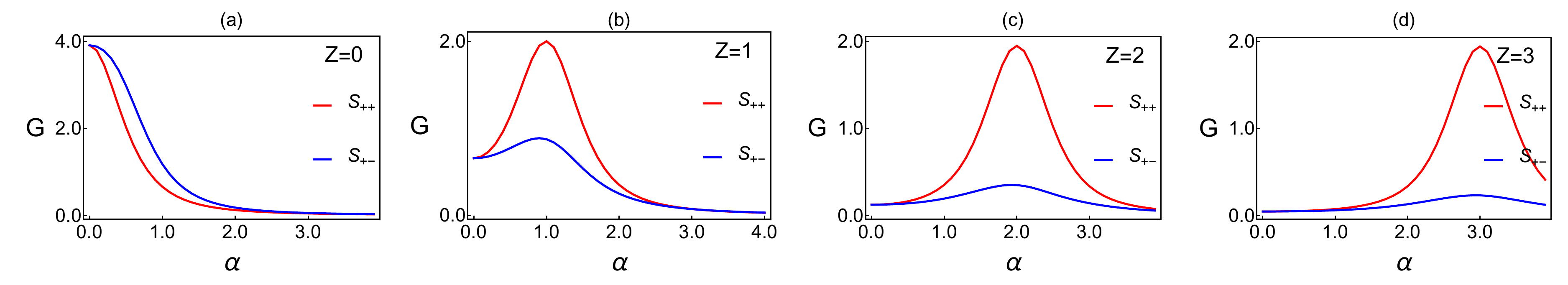}
        \caption{Conductance $G$ (in units of $\tfrac{2e^2}{h}$) as a function of $\alpha$ at $T=10K$ for different interband coupling strengths($Z$). Panels (a), (b), (c), and (d) correspond 
to $Z = 0$, $1$, $2$, and $3$, respectively, showing the conductance spectra for the 
$S_{++}$ pairing symmetry (red) and the $S_{+-}$ pairing symmetry (blue).}
        \label{cona}
    \end{figure}

\end{widetext}

\subsection{Finite temperature quantum noise}
At finite temperature, the total quantum noise $Q$ (in units of 
$\frac{4e^{2}k_{B}T}{h}$) is shown in Fig.~\ref{qnoisez} as a function of barrier strength $Z$ for different values of the interband coupling parameter $\alpha$. Fig.~\ref{qnoisea} shows the corresponding variation of the quantum noise as a function of $\alpha$ for different values of $Z$. 

For vanishing interband coupling ($\alpha=0$), the noise profiles corresponding to the $S_{++}$ and $S_{+-}$ pairing symmetries completely overlap for all values of the barrier strength $Z$. In this limit, both pairing symmetries exhibit a large noise value in the transparent regime ($Z\approx0$), followed by a monotonic suppression of the noise as the junction gradually enters the tunneling regime. The identical behaviour of the two pairing symmetries at $\alpha=0$ indicates that, in the absence of interband coupling, the phase difference between the superconducting gaps does not influence the transport noise properties.

However, once a finite interband coupling is introduced, the behaviour of the two pairing symmetries becomes qualitatively different. For $\alpha=1$, the $S_{++}$ pairing symmetry develops a pronounced peak structure around $Z\approx1$, whereas the $S_{+-}$ state shows only a much weaker enhancement in the same region. As the interband coupling strength is increased further to $\alpha=2$, the peak corresponding to the $S_{++}$ state becomes significantly sharper and larger, while the $S_{+-}$ branch exhibits only a weak and broad maximum. This distinction persists even for stronger coupling ($\alpha=3$), where the maximum of the $S_{++}$ response shifts towards larger values of the barrier strength, whereas the $S_{+-}$ state continues to show only a comparatively suppressed response.

\begin{widetext}

\begin{figure}[H]
        \centering
        \includegraphics[width=1.0\linewidth]{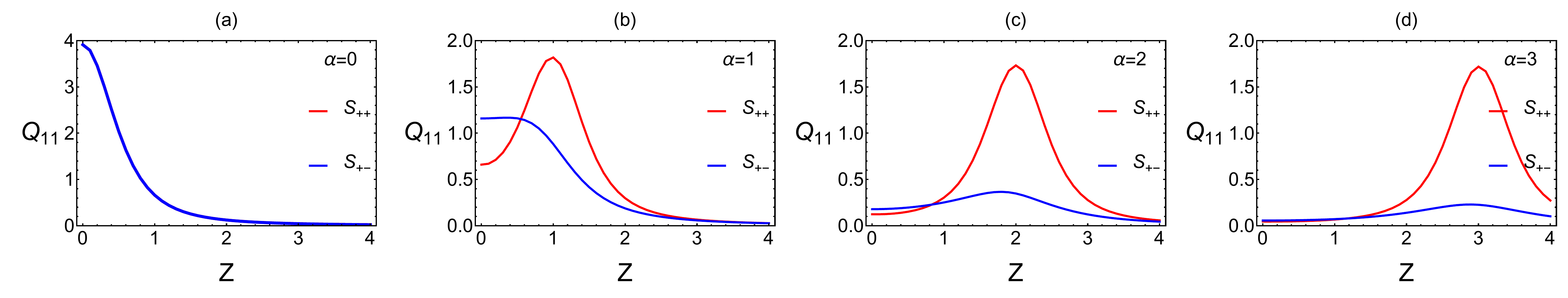}
        \caption{Finite temperature quantum noise \(Q_{\mathrm{}}\) (in units of \(\frac{4e^{2}}{h}k_BT\)) at $eV=0.10\Delta_1$, $T_1 = T_2 = 10K$ as a function of the barrier strength \(Z\) for different interband coupling strengths($\alpha$). Panels (a), (b), (c), and (d) correspond to \(\alpha = 0,\,1,\,2,\,3\), respectively, showing the behaviour of the \(S_{++}\) state (red) and the \(S_{+-}\) state (blue), with $\Delta_2=1.5\Delta_1$}
        \label{qnoisez}
    \end{figure}
\end{widetext}

A similar trend is observed in Fig.~\ref{qnoisea}, where the quantum noise is plotted as a function of interband coupling strength $\alpha$ for different barrier strengths $Z$. For a transparent junction ($Z=0$), the noise corresponding to the $S_{++}$ and $S_{+-}$ pairing symmetries remains nearly identical over the entire range of $\alpha$. In this regime, both pairing states exhibit a monotonic decrease in the noise with increasing interband coupling strength. However, once the barrier strength becomes finite, a clear distinction between the two pairing symmetries emerges. 

For $Z=1$, the $S_{++}$ pairing symmetry develops a pronounced peak at intermediate values of $\alpha$, while the $S_{+-}$ state exhibits only a smaller and broader maximum. As the barrier strength increases further ($Z=2$ and $Z=3$), the peak in the $S_{++}$ branch becomes increasingly sharp and prominent, with its position shifting towards larger $\alpha$. In contrast, the $S_{+-}$ state continues to display only a weak enhancement with significantly reduced amplitude. Thus, the separation between the two pairing symmetries becomes progressively more visible in the tunneling regime.

These results demonstrate that finite-temperature quantum noise can distinguish between the $S_{++}$ and $S_{+-}$ pairing symmetries through their qualitatively different responses to the interband coupling and barrier strength. In particular, the emergence of strong resonant peak structures in the $S_{++}$ state, contrasted with the comparatively suppressed response of the $S_{+-}$ state, provides a useful signature of the underlying superconducting pairing symmetry. Nevertheless, the distinction between the two pairing states is still not extremely sharp at finite temperature due to thermal broadening effects. Therefore, although quantum noise provides important information regarding the pairing symmetry, a more sensitive probe may be required for an unambiguous distinction between the $S_{++}$ and $S_{+-}$ superconducting states.

\begin{widetext}

\begin{figure}[H]
        \centering
        \includegraphics[width=1.0\linewidth]{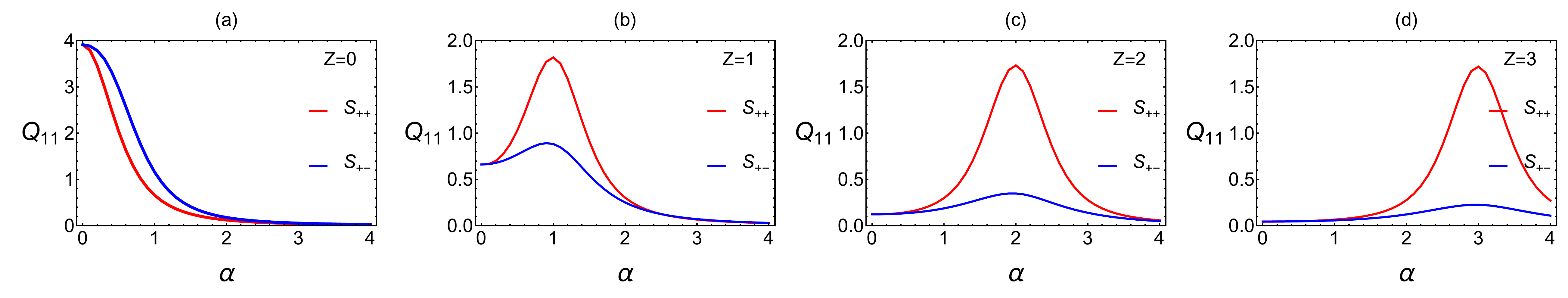}
        \caption{Finite temperature \(Q_{\mathrm{}}\) (in units of \(\frac{4e^{2}}{h}k_BT\)) at $eV=0.10\Delta_1$ at $T=10K$ as a function of the barrier strength \(\alpha\) for different interband coupling strengths($\alpha$). Panels (a), (b), (c), and (d) correspond to \(Z = 0,\,1,\,2,\,3\), respectively, showing the behaviour of the \(S_{++}\) state (red) and the \(S_{+-}\) state (blue), with $\Delta_2=1.5\Delta_1$}
        \label{qnoisea}
    \end{figure}
\end{widetext}

\subsection{Zero temperature quantum shot noise}

% \subsubsection{Zero temperature}

Quantum shot noise \( Q_{11}^{\mathrm{sh}} \) (in units of \(\frac{4e^{2}}{h}eV\)) at zero temperature is depicted as a function of the barrier strength \( Z \) for several interband coupling strengths \(\alpha\), see Fig.~\ref{fig5}. Shot noise is calculated at \( eV = 0.99\Delta_1 \) for two distinct superconducting pairing symmetries, \( S_{++} \) (red curves) and \( S_{+-} \) (blue curves).
% At this bias voltage, where $\Delta_1 < 0.99\Delta_1 < \Delta_2$, the energy window lies between the two superconducting gaps
% .
In this regime, the electron-hole asymmetry is maximized, leading to a pronounced effect on the quantum shot noise. For vanishing coupling (\( \alpha = 0 \)), both \( S_{++} \) and \( S_{+-} \) pairing symmetries produce identical shot noise profiles. Both feature a single peak at low barrier strength, which rapidly decays as \( Z \) increases, indicating limited sensitivity to the underlying pairing state in the absence of interband mixing. At $\alpha = 1$, a pronounced dip is observed at $Z \approx 1$ for the 
$S_{++}$ pairing symmetry, whereas the $S_{+-}$ state exhibits a peak 
in the same regime. With moderate interband coupling (\( \alpha = 2 \)), a pronounced difference emerges. The shot noise for the \( S_{++} \) symmetry shows twin peak like structure a dip in the intermediate barrier strength ($Z \approx 2$) regime. In contrast, the \( S_{+-} \) state shows a single peak around $Z=2$. For stronger interband coupling ($\alpha = 3$), the overall behaviour closely follows the $\alpha = 2$ case, the $S_{++}$ state develops a dip, while the $S_{+-}$ curve gives a peak. The main difference is a systematic shift of the $S_{++}$ minimum from $Z \approx 2$ (for $\alpha = 2$) to $Z \approx 3$ at $\alpha = 3$. 
% This indicates that increasing $\alpha$ 
% primarily displaces the interference-induced destructive-noise feature toward 
% larger barrier strengths.

These panels collectively demonstrate that shot noise, as a function of barrier strength and tunable interband coupling, provides a robust probe for discriminating \( S_{++} \) from \( S_{+-} \) pairing states in Iron Pnictide superconductors. The emergence and evolution of the twin-peak structure in \( S_{++} \) at higher \( \alpha \) values serves as a clear experimental fingerprint, making noise spectroscopy across tunable barriers a powerful diagnostic for investigating Iron Pnictide order parameter symmetry.

\begin{widetext}

\begin{figure}[H]
        \centering
        \includegraphics[width=1.0\linewidth]{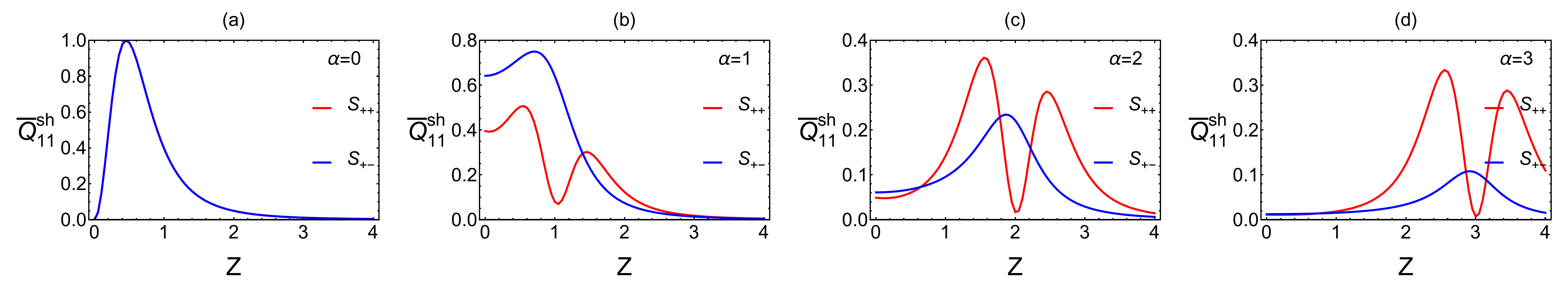}
        \caption{Zero temperature quantum shot noise \(Q_{\mathrm{sh}}\) (in units of \(\frac{4e^{2}}{h}eV\)) at $eV=0.10\Delta_1$ as a function of the barrier strength \(Z\) for different interband coupling strengths($\alpha$) with $T_1=T_2=0K$. Panels (a), (b), (c), and (d) correspond to \(\alpha = 0,\,1,\,2,\,3\), respectively, showing the behaviour of the \(S_{++}\) (red) and the \(S_{+-}\) pairing symmetry (blue), with $\Delta_2=1.5\Delta_1$}
        \label{fig5}
    \end{figure}
\end{widetext}

Fig.~\ref{fig6} shows the quantum shot noise as a function of the 
interband coupling strength $\alpha$ for different barrier strengths. 
For a transparent interface ($Z=0$), the $S_{++}$ and $S_{+-}$ pairing 
symmetries exhibit nearly identical behaviour, characterized by a broad 
maximum at low $\alpha$ followed by a gradual decay, indicating that 
interband coupling alone cannot distinguish the pairing states in the 
absence of a barrier. 

At $Z=1$, a clear contrast emerges: the $S_{++}$ symmetry develops a dip 
around $\alpha \approx 1$, whereas the $S_{+-}$ state shows a peak in the 
same regime. For larger barrier strengths ($Z=2$ and $Z=3$), this 
symmetry-dependent behaviour becomes more pronounced. The $S_{++}$ state 
exhibits a dip (leading to a twin-peak structure), while the $S_{+-}$ 
state shows a single peak near $\alpha \approx Z$, resulting in a 
substantial separation between the two responses.

\begin{widetext}

\begin{figure}[H]
        \centering
        \includegraphics[width=1.0\linewidth]{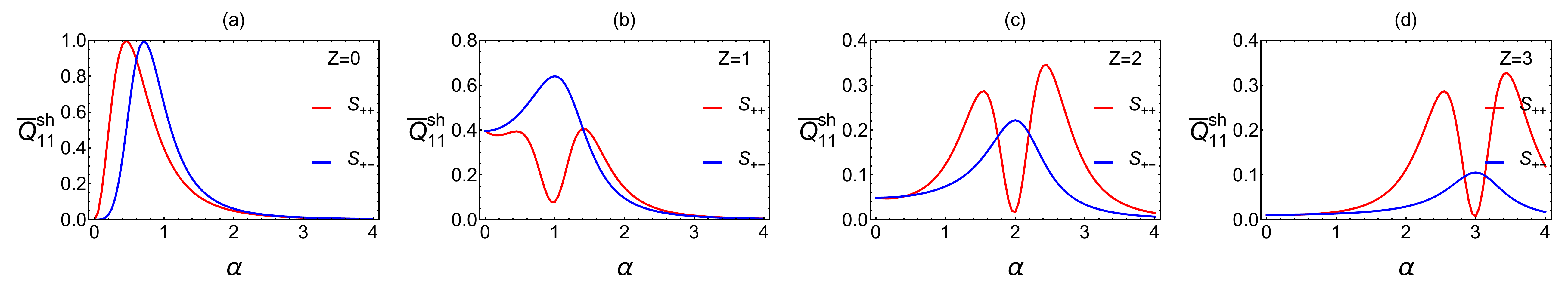}
        \caption{Zero temperature quantum shot noise \(Q_{\mathrm{sh}}\) (in units of \(\frac{4e^{2}}{h}eV\)) at $eV=0.10\Delta_1$ as a function of the interband coupling strength \(\alpha\) for different barrier strengts($Z$) with $T_1=T_2=0K$. Panels (a), (b), (c), and (d) correspond to \(Z = 0,\,1,\,2,\,3\), respectively, showing the behaviour of the \(S_{++}\) pairing symmetry (red) and the \(S_{+-}\) pairing symmetry (blue), with $\Delta_2=1.5\Delta_1$}
        \label{fig6}
    \end{figure}
\end{widetext}

% \clearpage
\subsection{$\Delta_T$ noise}
The dependence of \(\Delta_T\) noise (in units of \(\frac{4e^2}{h}k_BT\)) on barrier strength \(Z\) is systematically shown in Fig.~\ref{fig13} at temperature \(T_{\mathrm{1}} = 11\,\mathrm{K}\), \(T_{\mathrm{2}} = 9\,\mathrm{K}\) and thermal bias \(\Delta T = 1\,\mathrm{K}\) and varying interband coupling parameter \(\alpha\), focusing on the two distinct pairing symmetries: \(S_{++}\) (red) and \(S_{+-}\) (blue). 

For vanishing interband coupling ($\alpha = 0$), both pairing states produce essentially identical noise profiles. $\Delta_T$ noise exhibits a single broad maximum at low barrier strengths and decreases monotonically with increasing $Z$. In this case, the \(\Delta_T\) noise is dominated by single-band processes, and the absence of interband mixing leads to no difference between the two pairing symmetries. At $\alpha = 0$, the system effectively reduces to a single-band problem, and the resulting $\Delta_T$ noise exhibits behaviour qualitatively similar to that of a normal metal--insulator--superconductor junction, as reported in Ref.~\cite{AndreevDTnoise}. At $\alpha=1$ $S_{++}$ shows a dip whereas $S_{+-}$ shows a peak. For moderate interband coupling ($\alpha = 2$), the two pairing symmetries begin to show markedly different signatures. The $S_{++}$ state develops a dip arounf $Z\approx\alpha$, whereas the $S_{+-}$ state shows peak. This contrast arises from the onset of interband interference, and it can distinguish between two pairing symmetries. With stronger interband coupling ($\alpha = 3$), the qualitative trends observed for $\alpha = 2$ persist but become more pronounced. The dip in $S_{++}$ shifts slightly toward higher barrier strengths and increases in amplitude, while the $S_{+-}$ curve shows apeak around $Z\approx 3$.

% \newpage
\begin{widetext}

    \begin{figure}[H]
        \centering
        \includegraphics[width=1.0\linewidth]{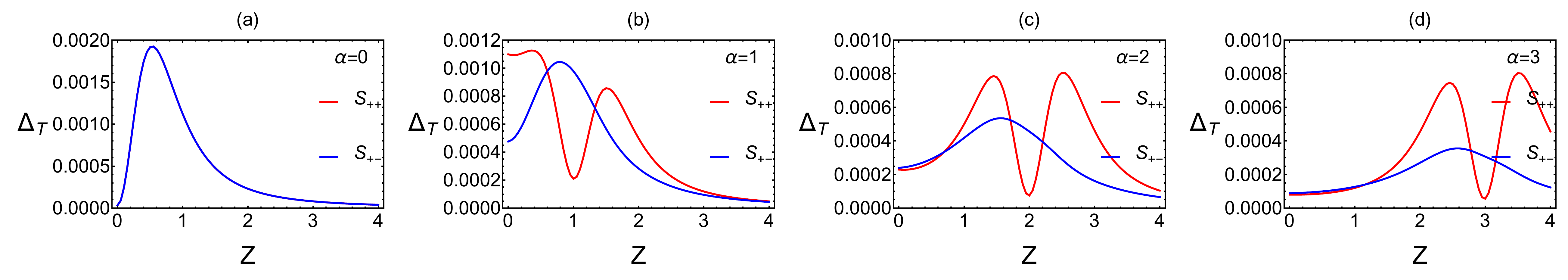}
        \caption{$\Delta_T$ noise (in units of \(\frac{4e^{2}}{h}k_{B}T\)) as a function of the barrier strength \(Z\) for different interband coupling strengths, at $T_1 = 11K$ and $T_2 = 9$ with $\Delta T = 1K$. Panels (a), (b), and (c) correspond to \(\alpha = 0,\,1,\,2,\,3\), respectively, showing the behavior of the \(S_{++}\) state (red) and the \(S_{+-}\) state (blue).}
        \label{fig13}
    \end{figure}
\end{widetext}

The dependence of the $\Delta_T$ noise (in units of $\frac{4e^2}{h}k_BT$) on the interband coupling strength $\alpha$ is shown in 
Fig.~\ref{fig14} at temperatures $T_1 = 11\,\mathrm{K}$ and 
$T_2 = 9\,\mathrm{K}$ under a fixed thermal bias $\Delta T = 1\,\mathrm{K}$, 
for different barrier strengths $Z$. The responses of the two pairing 
symmetries, $S_{++}$ (red) and $S_{+-}$ (blue), are compared. For a transparent interface both pairing symmetry have same $\Delta_T$ noise in the limit $\alpha=0$, as interband coupling strength increases $S_{++}$ pairing symmetry shows a peak and then decreases goes to zero in the high inetrband coupling strength limit. $S_{+-}$ pairing summetry shows two small peak and then decreaxes such vanishes in the higher $\alpha$ regimes. As the  barrier strength is increased to moderate values ($Z=2$), clear distinctions between the two pairing symmetries emerge. The $S_{++}$ 
state develops a suppression in the $\Delta_T$ noise around $\alpha \approx Z$, whereas the $S_{+-}$ pairing exhibits a peak in the same parameter regime. This contrasting behaviour signals the onset of interband-interference effects, which give rise to phase-sensitive $\Delta_T$ noise responses.

% For stronger barrier strength ($Z = 3$), the qualitative trends observed 
% at moderate coupling persist but become more pronounced. The suppression in the 
% $S_{++}$ branch deepens and shifts toward larger values of $\alpha$, while the 
% $S_{+-}$ response displays a pronounced peak near $\alpha \approx Z$.

\begin{widetext}

    \begin{figure}[H]
        \centering
        \includegraphics[width=1.0\linewidth]{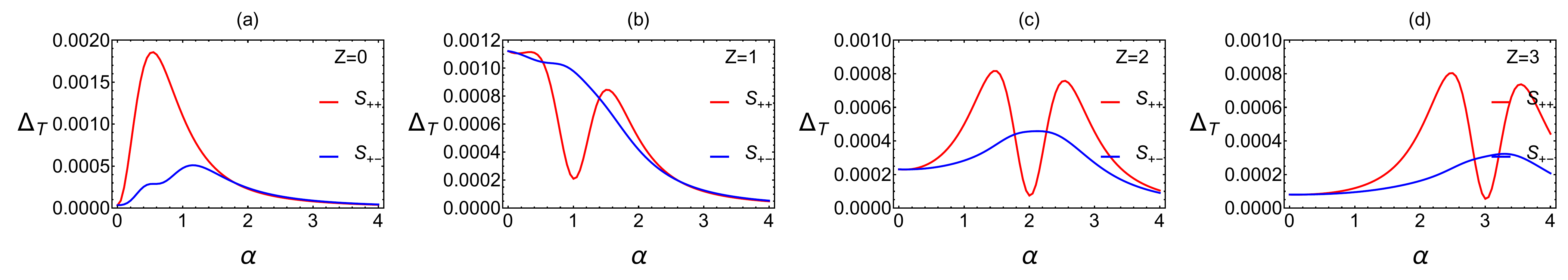}
        \caption{$\Delta_T$ noise (in units of \(\frac{4e^{2}}{h}k_{B}T\)) as a function of the interband coupling strength \(Z\) for different barrier strengths at $T_1 = 11K$ and $T_2 = 9$ with $\Delta T = 1K$. Panels (a), (b), and (c) correspond to \(Z = 0,\,1,\,2,\,3\), respectively, showing the behavior of the \(S_{++}\) state (red) and the \(S_{+-}\) state (blue).}
        \label{fig14}
    \end{figure}
\end{widetext}

\clearpage
\section{Analysis}
\label{analysis}

\begin{table*}[t]
\centering
\resizebox{0.95\textwidth}{!}{%
\begin{tabular}{|c|c|c|c|}
\hline
\textbf{Property} & $\mathbf{S}_{++}$ & $\mathbf{S}_{\pm}$ & Remarks \\
\hline

Conductance 
& Large peak at $Z\approx\alpha$ 
& Small peak at $Z\approx\alpha$ 
& Less effective\\
\hline

Finite temperature quantum noise 
& \begin{tabular}[c]{@{}c@{}}Large peak structure around $Z \approx \alpha$\end{tabular}
& Small peak at $Z\approx\alpha$
& Less effective\\
\hline

Zero temperature quantum shot noise 
& \begin{tabular}[c]{@{}c@{}}Shows \textbf{twin peak} structure\\ and exhibits a dip around $Z \approx \alpha$\end{tabular}
& Exhibits a \textbf{single peak} around $Z = \alpha$
& Very effective\\
\hline

$\Delta_T$ noise 
& \begin{tabular}[c]{@{}c@{}}Shows \textbf{twin peak} structure and exhibits a dip\\ at $Z=\alpha$ (vs $Z$), wherein $S_{+-}$ shows a peak\end{tabular}
& \begin{tabular}[c]{@{}c@{}}Shows a \textbf{single peak} at $Z=\alpha$ (vs $Z$),\\ wherein $S_{++}$ shows a dip\end{tabular}
& Very effective\\
\hline

\end{tabular}%
}
\caption{Comparison of key transport and noise properties for the $S_{++}$ and $S_{+-}$ pairing states.}
\label{tab}
\end{table*}

Table~\ref{tab} provides a comparative overview of several key transport properties for the \(S_{++}\) and \(S_{\pm}\) pairing states in a multiband superconducting system. The conductance is observed to be relatively symmetric in the \(S_{++}\) state, indicating uniform charge transport across the junction, whereas the \(S_{\pm}\) state exhibits a pronounced asymmetry, reflecting the impact of the sign-changing superconducting order parameter on quasiparticle transmission. This asymmetry arises due to the interference between intra- and interband scattering processes, which is more prominent in the \(S_{\pm}\) pairing symmetry.

{The common physical origin of all the transport and noise signatures presented in this work lies in the competition between interface scattering, governed by the barrier strength $Z$, and interband scattering, characterized by $\alpha$, together with the phase relation between the superconducting order parameters. The barrier $Z$ controls the probability of quasiparticles entering the superconducting region, whereas $\alpha$ mixes quasiparticle states belonging to different superconducting bands. Since the two pairing symmetries differ only in their relative phase, the interband scattering process causes qualitatively different quantum interference between the scattering amplitudes associated with the two bands. For the $S_{++}$ pairing symmetry, where the two superconducting gaps are in phase, the interband scattering contributes constructively, leading to enhanced scattering that manifest themselves as pronounced twin-peak structures in the quantum shot noise and $\Delta_T$ noise. In contrast, for the $S_{+-}$ pairing symmetry, the $\pi$ phase difference changes the interference between quasiparticles scattered from the two superconducting bands, leading to a single-peak instead of the twin-peaks observed for the $S_{++}$ pairing symmetry. Consequently, although all the transport quantities are calculated from the same scattering amplitudes, each observable combines these amplitudes in a different manner. Therefore, the conductance, finite-temperature quantum noise, zero-temperature quantum shot noise, and $\Delta_T$ noise exhibit different sensitivities to the underlying pairing symmetry. This enhanced sensitivity enables these noise observables to capture qualitative changes in the interference pattern, making them particularly effective probes for distinguishing between the $S_{++}$ and $S_{+-}$ pairing symmetries.}

The behaviour of the zero temperature quantum shot noise differs markedly for the two pairing symmetries. For the $S_{++}$ pairing symmetry, the shot noise exhibits a clear 
dip in the vicinity of $Z \approx \alpha$, whereas the $S_{+-}$ pairing symmetry 
shows a peak in the same parameter regime. This behaviour can be understood 
with the help of Fig.~\ref{noiseapp}, where we plot the individual contributions 
to the shot noise arising from the coefficients $c_1$, $c_2$, and $c_3$, 
corresponding to the terms $(f_{1h}-f_{1e})^2$, $(f_{1e}-f_{2e})^2$, and 
$(f_{1h}-f_{2e})^2$ in the noise expression, respectively.

The coefficients $c_2$ and $c_3$ exhibit identical behaviour (see Fig.~\ref{noiseapp} in Appendix ~\ref{quantum noise appendix}). 
For the $S_{++}$ pairing symmetry, both coefficients display a twin peak structure, 
with a dip appearing around $Z \approx \alpha$. A similar suppression is also present for the $S_{+-}$ pairing symmetry near the same value of $Z$. However, the overall shot noise response is governed by the relative magnitudes of these coefficients 
and their interplay with $c_1$. For the $S_{++}$ pairing symmetry, $c_1$ remains significantly smaller than $c_2$ 
and $c_3$ in the vicinity of $Z \approx \alpha$. Consequently, the dip in $c_2$ and 
$c_3$ directly manifests itself in the shot noise, giving rise to a twin peak structure. In contrast, for the $S_{+-}$ pairing symmetry, $c_1$ exhibits a pronounced peak around $Z \approx \alpha$ with a magnitude comparable to that of $c_2$ and $c_3$, this contribution compensates the dip produced by $c_2$ and $c_3$, thereby eliminating the twin peak structure and resulting in a single prominent peak in the total shot noise.

The finite-temperature quantum noise is evaluated at $eV = 0.10\Delta_1$, chosen such that the applied voltage bias remains within the linear response regime. In this regime, the voltage scale is sufficiently small compared to the characteristic superconducting energy scale, and therefore the total quantum noise is predominantly governed by the {equilibrium noise} contribution, while the purely nonequilibrium shot-noise component remains comparatively weak. Consequently, the qualitative behavior of the finite-temperature quantum noise closely resembles that of the conductance. In particular, the $S_{++}$ pairing symmetry exhibits a pronounced peak structure, whereas the $S_{+-}$ state also shows a single-peak profile, but with significantly reduced amplitude. This demonstrates that even at finite temperature, the dominant features of the quantum noise in the linear response regime are primarily controlled by thermal fluctuations. As a result, the finite-temperature noise retains the essential qualitative signatures of the underlying superconducting pairing symmetry, although thermal broadening partially suppresses the contrast between the $S_{++}$ and $S_{+-}$ states.

Table \ref{tab} demonstrates that conductance, finite temperature quantum noise, zero temperature quantum shot noise, 
and \(\Delta_T\) noise collectively form a robust set of observables capable of distinguishing 
\(S_{++}\) and \(S_{\pm}\) pairing symmetries in multiband superconductors. 
Together, they provide a coherent and experimentally accessible framework for identifying 
unconventional interband phase structures. Our analysis shows that distinguishing multiband pairing states requires a combination of transport, noise probes rather than any single observable. Because each quantity is affected differently by the interband phase structure, their collective behavior produces a clear and symmetry-dependent signature. While Ref.~\cite{PhysRevB.108.L100511} has proposed thermoelectric coefficients as probes of pairing symmetry; these signatures are primarily quantitative. In contrast, we demonstrate that noise observables particularly $\Delta_T$ noise and quantum shot noise exhibit qualitatively distinct spectral structures for $S_{++}$ (twin peak) vs. $S_{+-}$
(single peak) pairings, providing a far more definitive transport fingerprint.

% This combined approach complements and extends the thermoelectric criteria of Ref.~\cite{PhysRevB.108.L100511}, offering a more complete set of experimental indicators for separating the $S_{++}$ and $S_{+-}$ pairing states.

\section{Experimental realization and Conclusion}

% \subsection{Fabrication of the N-I-IP junction}

The proposed signatures can be tested in a normal metal–insulator–Iron
Pnictide superconductor (N-I-IP) junction. Such junctions can be
fabricated experimentally by depositing a thin insulating barrier
between a normal metal electrode (e.g., Au, Ag, or Pt) and an
iron–pnictide superconductor. The insulating layer can be realized
either by controlled oxidation of the surface or by inserting a thin
oxide barrier during the deposition process, thereby forming a tunnel
junction with adjustable interface transparency.
Alternatively, the junction can also be realized through point-contact
spectroscopy, where a normal metal tip is pressed against the
superconducting surface to form a nanoscale contact
\cite{PhysRevB.27.112}. Iron–pnictide superconductors such as LiFeAs
provide suitable platforms owing to their multiband electronic
structure and the availability of high-quality single crystals and
thin films \cite{Borisenko2010LiFeAs}. These experimental configurations
allow controlled investigation of charge transport, noise properties across the interface.

Typical experiments on Iron Pnictide junctions are performed at
temperatures of a few Kelvin~\cite{Nag2016}, well below the superconducting
transition temperature ($T_c \sim 15$--$40$ K depending on the
compound). In this regime the superconducting gaps are of the order of a few meV, which lies well within the resolution of modern tunneling spectroscopy and noise measurement techniques.

% \subsection{Control of the barrier strength $Z$}

% Within the Blonder--Tinkham--Klapwijk (BTK) formalism, the parameter
% $Z$ characterizes the strength of the interfacial potential barrier
% between the normal metal and the superconductor and therefore
% determines the transparency of the junction \cite{BTK}. Small values
% of $Z$ correspond to highly transparent interfaces where Andreev
% reflection dominates the transport, whereas large values of $Z$
% describe the tunneling regime.

Experimentally, the effective junction barrier strength can be controlled by
modifying the interface transparency. In tunnel junctions this can be
achieved by varying the thickness or oxidation level of the insulating
layer at the interface. In point-contact spectroscopy the transparency
can be tuned mechanically by adjusting the pressure of the metallic
tip on the superconducting surface, thereby modifying the effective
contact area and scattering at the interface
\cite{PhysRevB.27.112}. By varying these interfacial conditions,
a wide range of transport regimes corresponding to different values
of $Z$ can be accessed experimentally.

% \subsection{Origin of the interband coupling parameter $\alpha$}

The parameter $\alpha$ in the theoretical model represents the
effective coupling between the superconducting bands. In iron--pnictide
superconductors, such interband coupling naturally arises from the
multiorbital electronic structure of say the FeAs layers and from
interband scattering processes mediated by impurities or orbital
hybridization \cite{PhysRevLett.101.057003,Chubukov}. The interband coupling strength at the interface is theoretically estimated to lie in the range $0\text{--}3$~\cite{PhysRevB.80.144507,Liu2014Interplay}, and all calculations in this work are performed within this regime. These processes mix the
quasiparticle states belonging to different Fermi surface sheets,
leading to effective interband pairing interactions and scattering
between bands. Experimentally, the strength of such interband processes can vary
depending on material composition, doping level, and disorder.
Consequently, different iron--pnictide compounds~\cite{PhysRevLett.101.087004} or sample conditions
can effectively correspond to different values of the interband
coupling parameter $\alpha$ in the theoretical description.

% \subsection{Measurement techniques}

% \subsubsection{Conductance}

The differential conductance of the N-I-IP junction can be measured
using standard tunneling spectroscopy techniques. A small AC
modulation voltage is superimposed on the applied DC bias and the
resulting current response is detected using lock--in amplification.
The differential conductance $dI/dV$ is then obtained directly from
the measured signal. Such measurements have been widely used to probe
Andreev reflection and superconducting gap structures in normal
metal–superconductor junctions \cite{PhysRevB.27.112}.

% \subsubsection{Finite-temperature quantum noise}

At finite temperature the total quantum noise in the junction contains
contributions from both thermal fluctuations and shot noise.
Experimentally, this noise can be measured by detecting the current
or voltage fluctuations across the junction using cryogenic low-noise
amplifiers and spectrum analyzers~\cite{Jehl2000}. The noise spectral density is
obtained from the power spectrum of the measured fluctuations.
Such measurements have been widely employed in mesoscopic transport
experiments to probe quasiparticle transport and superconducting
correlations \cite{Blanter2000}. At sufficiently low temperatures the noise is dominated by quantum
shot noise originating from the discrete nature of charge transport.
Shot noise can be measured by analyzing current fluctuations under an
applied bias voltage using sensitive cross-correlation noise
detection techniques. Shot noise measurements have been successfully
used to study Andreev reflection and quasiparticle transport in
superconducting junctions \cite{deJong1994ShotNoise,Spietz2003ShotNoise}.

% \subsubsection{Thermoelectric measurements}

The $\Delta_T$ noise can be measured by maintaining a temperature
difference across the junction {with zero thermovoltage}, where the net current through the device vanishes~\cite{Lumbroso2018}. Under this
zero-current condition the quantum shot noise correspond
to the $\Delta_T$ noise component. Such measurements require sensitive
noise detection techniques combined with precise control of the
temperature gradient across the junction, similar to setups used in
mesoscopic noise spectroscopy experiments.

It is worth noting that resonance phenomena arising from the interplay of interface parameters have also been reported in spinful superconducting junctions with spin-orbit coupling~\cite{PhysRevB.101.014515}. Although the underlying microscopic model is fundamentally different from the present two-band Iron Pnictide system, exploring a possible correspondence between the two approaches would be an interesting direction for future investigation.

% \section{Conclusion}

% \section{Conclusion}

In summary, we have examined the transport and noise characteristics of a normal metal–insulator–iron-pnictide junction to identify signatures of the superconducting pairing symmetry. While conventional conductance measurements provide useful information about quasiparticle transport, they are often insufficient to unambiguously distinguish between the $S_{++}$ and $S_{+-}$ states in multiband superconductors, as both pairing symmetries typically produce similar single-peak conductance spectra that mainly differ in magnitude. To overcome this limitation, we explored alternative probes that are more sensitive to interband coupling and electron–hole asymmetry. Our analysis shows that temperature-driven noise, $\Delta_T$ noise, provides a clear qualitative distinction between the two pairing symmetries, with the $S_{++}$ state exhibiting a twin-peak structure and the $S_{+-}$ state displaying a single-peak profile. Similar symmetry-dependent behavior is observed in both zero-temperature quantum shot noise and finite-temperature quantum noise. Together, these complementary signatures demonstrate that combined noise spectroscopy measurements offer a robust and experimentally accessible framework for identifying the superconducting order parameter symmetry and probing interband coupling effects in iron-pnictide superconductors. Future work could extend the present analysis to multiterminal
iron-pnictide junctions, where nonlocal transport processes such as
crossed Andreev reflection become important. In such geometries,
nonlocal noise correlations may provide additional phase-sensitive
signatures of the pairing symmetry in multiband superconductors.

% \newpage
\section*{Appendix}
\label{appendix}
% The Appendix is divided into five sections. In Appendix \ref{s matrix} we discuss the method to get the scattering matrix elements. In Appendix \ref{quantum noise appendix} we discuss the calculation of quantum shot noise. In Appendix \ref{thermovoltage}, we calculate the thermovoltage and Seebeck coefficient.
\appendix
\begin{widetext}    
\section{S-matrix}
\label{s matrix}

The wave functions for an electron incident in band $1$ from the normal metal are given below, for N-I-IP junction as in Fig.~\ref{setup1},

\begin{subequations}\label{eq:psi}
\begin{align}
\psi_{N}(x) &=
\left(e^{ik_1 x} + r^{ee}_{11n} e^{-ik_1 x}\right)\phi^{N}_1
+ r^{he}_{11n} e^{ik_2 x}\phi^{N}_2
+ r^{ee}_{12n} e^{-ik_1 x}\phi^{N}_3
+ r^{he}_{12n} e^{ik_2 x}\phi^{N}_4, 
&& x<0, \label{eq:psi-1a} \\[6pt]
\psi_{IP}(x) &=
t^{ee}_{11n} e^{iq_{1e} x}\phi^{S}_1
+ t^{he}_{11n} e^{-iq_{1h} x}\phi^{S}_2
+ t^{ee}_{12n} e^{iq_{2e} x}\phi^{S}_3
+ t^{he}_{12n} e^{-iq_{2h} x}\phi^{S}_4, 
&& x>0. \label{eq:psi-1b}
\end{align}
\label{wavefunction}
\end{subequations}

\[
\text{with } \phi_1^N = \begin{pmatrix} 1 \\ 0 \\ 0 \\ 0 \end{pmatrix}, \quad
\phi_2^N = \begin{pmatrix} 0 \\ 1 \\ 0 \\ 0 \end{pmatrix}, \quad
\phi_3^N = \begin{pmatrix} 0 \\ 0 \\ 1 \\ 0 \end{pmatrix}, \quad
\phi_4^N = \begin{pmatrix} 0 \\ 0 \\ 0 \\ 1 \end{pmatrix}, \quad
\phi_1^S = \begin{pmatrix} u_1 \\ v_1 \\ 0 \\ 0 \end{pmatrix}, \quad
\phi_2^S = \begin{pmatrix} v_1 \\ u_1 \\ 0 \\ 0 \end{pmatrix} , \quad
\phi_3^S = \begin{pmatrix} 0 \\ 0 \\ u_2 \\ v_2 e^{-i\phi} \end{pmatrix}, \quad
\phi_4^S =  \begin{pmatrix} 0 \\ 0 \\ v_2 e^{i\phi} \\ u_2 \end{pmatrix}
\]

Boundary conditions are:

\begin{subequations}
\label{bcapp}
\begin{align}
\Psi_{N}\big|_{x = 0} &= \Psi_{IP}\big|_{x = 0}, \label{bcapp:a} \\[6pt]
\left. \frac{\partial}{\partial x} \left( \Psi_{IP} - \Psi_{N} \right) \right|_{x = 0}
&= 2m^* \left(
V \, \mathrm{diag}(\hat{1}, \hat{1})
+ \tilde{\alpha}_0 \, \mathrm{offdiag}(\hat{1}, \hat{1})
\right) \Psi_{N}\big|_{x = 0}. \label{bcapp:b}
\end{align}
\end{subequations}

Eq.~\eqref{wavefunction}, Eq.~\eqref{bcapp} gives the wave function related to an electron incident on band 1 of the normal metal. Since the probability current for a two-component wave function \cite{BTK} $\psi=(f,g)^T$is given as 
$\vec{J} = \frac{\hbar}{m} \bigl[ \operatorname{Im}(f^{*} \nabla f) - \operatorname{Im}(g^{*} \nabla g) \bigr]$

Probability $P=\frac{J_{scttered}}{J_{incidence}}$.
Following the above formula, the scattering amplitudes for the electron incident on band 1 are:
$a_{11n}=s_{11}^{he11}=r^{he}_{11n},\;
a_{12n}=s_{11}^{he21}=r^{he}_{12n},\;
b_{11n}=s_{11}^{ee11}=r^{ee}_{11n},\;
b_{12n}=s_{11}^{ee21}=r^{ee}_{12n},\;
c_{11n}=s_{21}^{ee11}=\sqrt{|u_1|^2-|v_1|^2}\,t^{ee}_{11n},\;
c_{12n}=s_{21}^{ee21}=\sqrt{|u_2|^2-|v_2|^2}\,t^{ee}_{12n},\;
d_{11n}=s_{21}^{he11}=\sqrt{|u_1|^2-|v_1|^2}\,t^{he}_{11n},\;
d_{12n}=s_{21}^{he21}=\sqrt{|u_2|^2-|v_2|^2}\,t^{he}_{12n}.$ The corresponding probabilities are $A_{11n}=|a_{11n}|^2,\; B_{11n}=|b_{11n}|^2,\;
A_{12n}=|a_{12n}|^2,\; B_{12n}=|b_{12n}|^2,\;
C_{11n}=|c_{11n}|^2,\; D_{11n}=|d_{11n}|^2,\;
C_{12n}=|c_{12n}|^2,\; D_{12n}=|d_{12n}|^2.$

% \begin{align*}
% a_{11n} &= s_{11}^{he11} = r^{he}_{11n}, &
% a_{12n} &= s_{11}^{he21} = r^{he}_{12n}, \\[6pt]
% b_{11n} &= s_{11}^{ee11} = r^{ee}_{11n}, &
% b_{12n} &= s_{11}^{ee21} = r^{ee}_{12n}, \\[6pt]
% c_{11n} &= s_{21}^{ee11} = \sqrt{|u_1|^2-|v_1|^2}\, t^{ee}_{11n}, &
% c_{12n} &= s_{21}^{ee21} = \sqrt{|u_2|^2-|v_2|^2}\, t^{ee}_{12n}, \\[6pt]
% d_{11n} &= s_{21}^{he11} = \sqrt{|u_1|^2-|v_1|^2}\, t^{he}_{11n}, &
% d_{12n} &= s_{21}^{he21} = \sqrt{|u_2|^2-|v_2|^2}\, t^{he}_{12n}.
% \end{align*}

% \begin{align*}
% A_{11n}=|a_{11n}|^2,\; B_{11n}=|b_{11n}|^2, \qquad
% A_{12n}=|a_{12n}|^2,\; B_{12n}=|b_{12n}|^2,\\
% C_{11n}=|c_{11n}|^2,\; D_{11n}=|d_{11n}|^2, \qquad
% C_{12n}=|c_{12n}|^2,\; D_{12n}=|d_{12n}|^2.
% \end{align*}
% \begin{align*}
% A_{11n} &= |a_{11n}|^2, & A_{12n} &= |a_{12n}|^2, \\
% B_{11n} &= |b_{11n}|^2, & B_{12n} &= |b_{12n}|^2, \\
% C_{11n} &= |c_{11n}|^2, & C_{12n} &= |c_{12n}|^2, \\
% D_{11n} &= |d_{11n}|^2, & D_{12n} &= |d_{12n}|^2.
% \end{align*}

Similarly, for an electron incident on band 2 from the normal metal, we can calculate the scattering amplitudes using the same boundary condition and the current conservation. The amplitudes are $a_{21n}=s_{11}^{he12}=r^{he}_{21n},\;
a_{22n}=s_{11}^{he22}=r^{he}_{22n},\;
b_{21n}=s_{11}^{ee12}=r^{ee}_{21n},\;
b_{22n}=s_{11}^{ee22}=r^{ee}_{22n},\;
c_{21n}=s_{21}^{ee12}=\sqrt{|u_1|^2-|v_1|^2}\,t^{ee}_{21n},\;
c_{22n}=s_{21}^{ee22}=\sqrt{|u_2|^2-|v_2|^2}\,t^{ee}_{22n},\;
d_{21n}=s_{21}^{he12}=\sqrt{|u_1|^2-|v_1|^2}\,t^{he}_{21n},\;
d_{22n}=s_{21}^{he22}=\sqrt{|u_2|^2-|v_2|^2}\,t^{he}_{22n}.$
 The corresponding probabilities are$A_{21n}=|a_{21n}|^2,\;
A_{22n}=|a_{22n}|^2,\;
B_{21n}=|b_{21n}|^2,\;
B_{22n}=|b_{22n}|^2,\;
C_{21n}=|c_{21n}|^2,\;
C_{22n}=|c_{22n}|^2,\;
D_{21n}=|d_{21n}|^2,\;
D_{22n}=|d_{22n}|^2.$

% \begin{align*}
% a_{21n} &= s_{11}^{he12} = r^{he}_{21n}, &
% a_{22n} &= s_{11}^{he22} = r^{he}_{22n}, \\[6pt]
% b_{21n} &= s_{11}^{ee12} = r^{ee}_{21n}, &
% b_{22n} &= s_{11}^{ee22} = r^{ee}_{22n}, \\[6pt]
% c_{21n} &= s_{21}^{ee12} = \sqrt{|u_1|^2-|v_1|^2}\; t^{ee}_{21n}, &
% c_{22n} &= s_{21}^{ee22} = \sqrt{|u_2|^2-|v_2|^2}\; t^{ee}_{22n}, \\[6pt]
% d_{21n} &= s_{21}^{he12} = \sqrt{|u_1|^2-|v_1|^2}\; t^{he}_{21n}, &
% d_{22n} &= s_{21}^{he22} = \sqrt{|u_2|^2-|v_2|^2}\; t^{he}_{22n}.
% \end{align*}

% The corresponding probabilities are:
% \begin{align*}
% A_{21n} &= |a_{21n}|^2, & A_{22n} &= |a_{22n}|^2, \\
% B_{21n} &= |b_{21n}|^2, & B_{22n} &= |b_{22n}|^2, \\
% C_{21n} &= |c_{21n}|^2, & C_{22n} &= |c_{22n}|^2, \\
% D_{21n} &= |d_{21n}|^2, & D_{22n} &= |d_{22n}|^2.
% \end{align*}

% --- Wavefunction ---
The wave function for a hole incident on band 1 of the normal metal is given by,
\begin{subequations}\label{eq:psi}
\begin{align}
\psi_{N}(x) &=
 r^{eh}_{11n} e^{-ik_1 x}\phi^{N}_1
+ \left(e^{-ik_2 x} + r^{hh}_{11n} e^{ik_2 x}\right)\phi^{N}_2
+ r^{eh}_{12n} e^{-ik_1 x}\phi^{N}_3
+ r^{hh}_{12n} e^{ik_2 x}\phi^{N}_4, 
&& x<0, \label{eq:psi-1a} \\[6pt]
\psi_{IP}(x) &=
t^{eh}_{11n} e^{iq_{1e} x}\phi^{S}_1
+ t^{hh}_{11n} e^{-iq_{1h} x}\phi^{S}_2
+ t^{eh}_{12n} e^{iq_{2e} x}\phi^{S}_3
+ t^{hh}_{12n} e^{-iq_{2h} x}\phi^{S}_4, 
&& x>0. \label{eq:psi-1b}
\end{align}
\end{subequations}
We can calculate the scattering amplitudes using the same boundary condition Eq.~\ref{bcapp} and the current conservation. The scattering amplitudes are
$a_{11nh}=s_{11}^{eh11}=r^{eh}_{11n},\;
a_{12nh}=s_{11}^{eh21}=r^{eh}_{12n},\;
b_{11nh}=s_{11}^{hh11}=r^{hh}_{11n},\;
b_{12nh}=s_{11}^{hh21}=r^{hh}_{12n},\;
c_{11nh}=s_{21}^{hh11}=\sqrt{|u_1|^2-|v_1|^2}\,t^{hh}_{11n},\;
c_{12nh}=s_{21}^{hh21}=\sqrt{|u_2|^2-|v_2|^2}\,t^{hh}_{12n},\;
d_{11nh}=s_{21}^{eh11}=\sqrt{|u_1|^2-|v_1|^2}\,t^{eh}_{11n},\;
d_{12nh}=s_{21}^{eh21}=\sqrt{|u_2|^2-|v_2|^2}\,t^{eh}_{12n}.$
The corresponding probabilities are
$A_{11nh}=|a_{11nh}|^2,\;
A_{12nh}=|a_{12nh}|^2,\;
B_{11nh}=|b_{11nh}|^2,\;
B_{12nh}=|b_{12nh}|^2,\;
C_{11nh}=|c_{11nh}|^2,\;
C_{12nh}=|c_{12nh}|^2,\;
D_{11nh}=|d_{11nh}|^2,\;
D_{12nh}=|d_{12nh}|^2.$

% --- Scattering amplitudes ---
% The scattering amplitudes are given by,
% \[
% \begin{aligned}
% a_{11nh} &= s_{11}^{eh11}=r^{eh}_{11n}, 
% &\quad a_{12nh} &= s_{11}^{eh21}=r^{eh}_{12n}, \\[6pt]
% b_{11nh} &= s_{11}^{hh11}=r^{hh}_{11n}, 
% &\quad b_{12nh} &= s_{11}^{hh21}=r^{hh}_{12n}, \\[6pt]
% c_{11nh} &= s_{21}^{hh11}=\sqrt{|u_1|^2-|v_1|^2}\; t^{hh}_{11n}, 
% &\quad c_{12nh} &= s_{21}^{hh21}=\sqrt{|u_2|^2-|v_2|^2}\; t^{hh}_{12n}, \\[6pt]
% d_{11nh} &= s_{21}^{eh11}=\sqrt{|u_1|^2-|v_1|^2}\; t^{eh}_{11n}, 
% &\quad d_{12nh} &= s_{21}^{eh21}=\sqrt{|u_2|^2-|v_2|^2}\; t^{eh}_{12n}.
% \end{aligned}
% \]

% % --- Probabilities ---
% The corresponding probabilities are given by,
% \[
% \begin{aligned}
% A_{11nh} &= |a_{11nh}|^2, &\quad A_{12nh} &= |a_{12nh}|^2, \\[6pt]
% B_{11nh} &= |b_{11nh}|^2, &\quad B_{12nh} &= |b_{12nh}|^2, \\[6pt]
% C_{11nh} &= |c_{11nh}|^2, &\quad C_{12nh} &= |c_{12nh}|^2, \\[6pt]
% D_{11nh} &= |d_{11nh}|^2, &\quad D_{12nh} &= |d_{12nh}|^2.
% \end{aligned}
% \]

% --- Wavefunction (band-2 hole incident) ---
The wave function for a hole incident on band 2 of the normal metal is given by,
\begin{subequations}\label{eq:psi2}
\begin{align}
\psi_{N}(x) &=
 r^{eh}_{21n} e^{-ik_1 x}\phi^{N}_1
+ r^{hh}_{21n} e^{ik_2 x}\phi^{N}_2
+ r^{eh}_{22n} e^{-ik_1 x}\phi^{N}_3
+ \left(e^{-ik_2 x} + r^{hh}_{22n} e^{ik_2 x}\right)\phi^{N}_4, 
&& x<0, \label{eq:psi2-1a} \\[6pt]
\psi_{IP}(x) &=
t^{eh}_{21n} e^{iq_{1e} x}\phi^{S}_1
+ t^{hh}_{21n} e^{-iq_{1h} x}\phi^{S}_2
+ t^{eh}_{22n} e^{iq_{2e} x}\phi^{S}_3
+ t^{hh}_{22n} e^{-iq_{2h} x}\phi^{S}_4, 
&& x>0. \label{eq:psi2-1b}
\end{align}
\end{subequations}

% --- Scattering amplitudes ---
We can calculate the scattering amplitudes using the same boundary condition Eq.~\ref{bcapp} and the current conservation. The scattering amplitudes are
$a_{21nh}=s_{11}^{eh12}=r^{eh}_{21n},\;
a_{22nh}=s_{11}^{eh22}=r^{eh}_{22n},\;
b_{21nh}=s_{11}^{hh12}=r^{hh}_{21n},\;
b_{22nh}=s_{11}^{hh22}=r^{hh}_{22n},\;
c_{21nh}=s_{21}^{hh12}=\sqrt{|u_1|^2-|v_1|^2}\,t^{hh}_{21n},\;
c_{22nh}=s_{21}^{hh22}=\sqrt{|u_2|^2-|v_2|^2}\,t^{hh}_{22n},\;
d_{21nh}=s_{21}^{eh12}=\sqrt{|u_1|^2-|v_1|^2}\,t^{eh}_{21n},\;
d_{22nh}=s_{21}^{eh22}=\sqrt{|u_2|^2-|v_2|^2}\,t^{eh}_{22n}.$
The corresponding probabilities are
$A_{21nh}=|a_{21nh}|^2,\;
A_{22nh}=|a_{22nh}|^2,\;
B_{21nh}=|b_{21nh}|^2,\;
B_{22nh}=|b_{22nh}|^2,\;
C_{21nh}=|c_{21nh}|^2,\;
C_{22nh}=|c_{22nh}|^2,\;
D_{21nh}=|d_{21nh}|^2,\;
D_{22nh}=|d_{22nh}|^2.$
% The scattering amplitudes are given by, 
% \[
% \begin{aligned}
% a_{21nh} &= s_{11}^{eh12}=r^{eh}_{21n}, 
% &\quad a_{22nh} &= s_{11}^{eh22}=r^{eh}_{22n}, \\[6pt]
% b_{21nh} &= s_{11}^{hh12}=r^{hh}_{21n}, 
% &\quad b_{22nh} &= s_{11}^{hh22}=r^{hh}_{22n}, \\[6pt]
% c_{21nh} &= s_{21}^{hh12}=\sqrt{|u_1|^2-|v_1|^2}\; t^{hh}_{21n}, 
% &\quad c_{22nh} &= s_{21}^{hh22}=\sqrt{|u_2|^2-|v_2|^2}\; t^{hh}_{22n}, \\[6pt]
% d_{21nh} &= s_{21}^{eh12}=\sqrt{|u_1|^2-|v_1|^2}\; t^{eh}_{21n}, 
% &\quad d_{22nh} &= s_{21}^{eh22}=\sqrt{|u_2|^2-|v_2|^2}\; t^{eh}_{22n}.
% \end{aligned}
% \]

% % --- Probabilities ---
% The corresponding probabilities are given by,
% \[
% \begin{aligned}
% A_{21nh} &= |a_{21nh}|^2, &\quad A_{22nh} &= |a_{22nh}|^2, \\[6pt]
% B_{21nh} &= |b_{21nh}|^2, &\quad B_{22nh} &= |b_{22nh}|^2, \\[6pt]
% C_{21nh} &= |c_{21nh}|^2, &\quad C_{22nh} &= |c_{22nh}|^2, \\[6pt]
% D_{21nh} &= |d_{21nh}|^2, &\quad D_{22nh} &= |d_{22nh}|^2.
% \end{aligned}
% \]

% ===============================
% Case 1: Electron incident on band 1
% ===============================
The wave function for an electron incident on band 1 of the iron-pnictide superconductor is
\begin{subequations}
\begin{align}
\psi_{N}(x) &=
t^{ee}_{11s} e^{-ik_1 x}\phi^{N}_1
+ t^{he}_{11s} e^{ik_2 x}\phi^{N}_2
+ t^{ee}_{12s} e^{-ik_1 x}\phi^{N}_3
+ t^{he}_{12s} e^{ik_2 x}\phi^{N}_4, 
&& x<0, \\[6pt]
\psi_{IP}(x) &=
\left(e^{-iq_{1e} x}+r^{ee}_{11s} e^{iq_{1e} x}\right)\phi^{S}_1
+ r^{he}_{11s} e^{-iq_{1h} x}\phi^{S}_2
+ r^{ee}_{12s} e^{iq_{2e} x}\phi^{S}_3
+ r^{he}_{12s} e^{-iq_{2h} x}\phi^{S}_4, 
&& x>0.
\end{align}
\end{subequations}

We can calculate the scattering amplitudes using the same boundary condition Eq.~\ref{bcapp} and the current conservation. 
The scattering amplitudes are
$a_{11s}=s_{22}^{he11}=r^{he}_{11s},\;
a_{12s}=s_{22}^{he21}=\sqrt{\tfrac{|u_2|^2-|v_2|^2}{|u_1|^2-|v_1|^2}}\,r^{he}_{12s},\;
b_{11s}=s_{22}^{ee11}=r^{ee}_{11s},\;
b_{12s}=s_{22}^{ee21}=\sqrt{\tfrac{|u_2|^2-|v_2|^2}{|u_1|^2-|v_1|^2}}\,r^{ee}_{12s},\;
c_{11s}=s_{12}^{ee11}=\tfrac{1}{\sqrt{|u_1|^2-|v_1|^2}}\,t^{ee}_{11s},\;
c_{12s}=s_{12}^{ee21}=\tfrac{1}{\sqrt{|u_1|^2-|v_1|^2}}\,t^{ee}_{12s},\;
d_{11s}=s_{12}^{he11}=\tfrac{1}{\sqrt{|u_1|^2-|v_1|^2}}\,t^{he}_{11s},\;
d_{12s}=s_{12}^{he21}=\tfrac{1}{\sqrt{|u_1|^2-|v_1|^2}}\,t^{he}_{12s}.$
The corresponding probabilities are
$A_{11s}=|a_{11s}|^2,\;
A_{12s}=|a_{12s}|^2,\;
B_{11s}=|b_{11s}|^2,\;
B_{12s}=|b_{12s}|^2,\;
C_{11s}=|c_{11s}|^2,\;
C_{12s}=|c_{12s}|^2,\;
D_{11s}=|d_{11s}|^2,\;
D_{12s}=|d_{12s}|^2.$

% The scattering amplitudes are
% \begin{align*}
% a_{11s}&= s_{22}^{he11}=r^{he}_{11s}, \quad
% &a_{12s}&= s_{22}^{he21}=\sqrt{\tfrac{|u_2|^2-|v_2|^2}{|u_1|^2-|v_1|^2}}\; r^{he}_{12s}, \\[4pt]
% b_{11s}&= s_{22}^{ee11}=r^{ee}_{11s}, \quad
% &b_{12s}&= s_{22}^{ee21}=\sqrt{\tfrac{|u_2|^2-|v_2|^2}{|u_1|^2-|v_1|^2}}\; r^{ee}_{12s}, \\[4pt]
% c_{11s}&= s_{12}^{ee11}=\tfrac{1}{\sqrt{|u_1|^2-|v_1|^2}}\; t^{ee}_{11s}, \quad
% &c_{12s}&= s_{12}^{ee21}=\tfrac{1}{\sqrt{|u_1|^2-|v_1|^2}}\; t^{ee}_{12s}, \\[4pt]
% d_{11s}&= s_{12}^{he11}=\tfrac{1}{\sqrt{|u_1|^2-|v_1|^2}}\; t^{he}_{11s}, \quad
% &d_{12s}&= s_{12}^{he21}=\tfrac{1}{\sqrt{|u_1|^2-|v_1|^2}}\; t^{he}_{12s}.
% \end{align*}

% The corresponding probabilities are
% \[
% A_{11s}=|a_{11s}|^2,\;\; A_{12s}=|a_{12s}|^2,\;\;
% B_{11s}=|b_{11s}|^2,\;\; B_{12s}=|b_{12s}|^2,\;\;
% C_{11s}=|c_{11s}|^2,\;\; C_{12s}=|c_{12s}|^2,\;\;
% D_{11s}=|d_{11s}|^2,\;\; D_{12s}=|d_{12s}|^2.
% \]

% ===============================
% Case 2: Electron incident on band 2
% ===============================
The wave function for an electron incident on band 2 is
\begin{subequations}
\begin{align}
\psi_{N}(x) &=
t^{ee}_{21s} e^{-ik_1 x}\phi^{N}_1
+ t^{he}_{21s} e^{ik_2 x}\phi^{N}_2
+ t^{ee}_{22s} e^{-ik_1 x}\phi^{N}_3
+ t^{he}_{22s} e^{ik_2 x}\phi^{N}_4, 
&& x<0, \\[6pt]
\psi_{IP}(x) &=
r^{ee}_{21s} e^{iq_{1e} x}\phi^{S}_1
+ r^{he}_{21s} e^{-iq_{1h} x}\phi^{S}_2
+ \left(e^{-iq_{2e} x}+r^{ee}_{22s} e^{iq_{2e} x}\right)\phi^{S}_3
+ r^{he}_{22s} e^{-iq_{2h} x}\phi^{S}_4, 
&& x>0.
\end{align}
\end{subequations}

We can calculate the scattering amplitudes using the same boundary condition Eq.~\ref{bcapp} and the current conservation. The scattering amplitudes are
$a_{21s}=\sqrt{\tfrac{|u_1|^2-|v_1|^2}{|u_2|^2-|v_2|^2}}\,r^{he}_{21s},\;
a_{22s}=r^{he}_{22s},\;
b_{21s}=\sqrt{\tfrac{|u_1|^2-|v_1|^2}{|u_2|^2-|v_2|^2}}\,r^{ee}_{21s},\;
b_{22s}=r^{ee}_{22s},\;
c_{21s}=\tfrac{1}{\sqrt{|u_2|^2-|v_2|^2}}\,t^{ee}_{21s},\;
c_{22s}=\tfrac{1}{\sqrt{|u_2|^2-|v_2|^2}}\,t^{ee}_{22s},\;
d_{21s}=\tfrac{1}{\sqrt{|u_2|^2-|v_2|^2}}\,t^{he}_{21s},\;
d_{22s}=\tfrac{1}{\sqrt{|u_2|^2-|v_2|^2}}\,t^{he}_{22s}.$
The corresponding probabilities are
$A_{21s}=|a_{21s}|^2,\;
A_{22s}=|a_{22s}|^2,\;
B_{21s}=|b_{21s}|^2,\;
B_{22s}=|b_{22s}|^2,\;
C_{21s}=|c_{21s}|^2,\;
C_{22s}=|c_{22s}|^2,\;
D_{21s}=|d_{21s}|^2,\;
D_{22s}=|d_{22s}|^2.$

% Scattering amplitudes:
% \begin{align*}
% a_{21s} &= \sqrt{\frac{|u_1|^2 - |v_1|^2}{|u_2|^2 - |v_2|^2}} \, r^{he}_{21s}, \quad
% &a_{22s} &= r^{he}_{22s}, \\[1mm]
% b_{21s} &= \sqrt{\frac{|u_1|^2 - |v_1|^2}{|u_2|^2 - |v_2|^2}} \, r^{ee}_{21s}, \quad
% &b_{22s} &= r^{ee}_{22s}, \\[1mm]
% c_{21s} &= \frac{1}{\sqrt{|u_2|^2 - |v_2|^2}} \, t^{ee}_{21s}, \quad
% &c_{22s} &= \frac{1}{\sqrt{|u_2|^2 - |v_2|^2}} \, t^{ee}_{22s}, \\[1mm]
% d_{21s} &= \frac{1}{\sqrt{|u_2|^2 - |v_2|^2}} \, t^{he}_{21s}, \quad
% &d_{22s} &= \frac{1}{\sqrt{|u_2|^2 - |v_2|^2}} \, t^{he}_{22s}.
% \end{align*}

% Probabilities:
% \[
% A_{21s}=|a_{21s}|^2,\;\; A_{22s}=|a_{22s}|^2,\;\;
% B_{21s}=|b_{21s}|^2,\;\; B_{22s}=|b_{22s}|^2,\;\;
% C_{21s}=|c_{21s}|^2,\;\; C_{22s}=|c_{22s}|^2,\;\;
% D_{21s}=|d_{21s}|^2,\;\; D_{22s}=|d_{22s}|^2.
% \]

% ===============================
% Case 3: Hole incident on band 1
% ===============================
The wave function for a hole incident on band 1 is
\begin{subequations}
\begin{align}
\psi_{N}(x) &=
 t^{eh}_{11sh} e^{-ik_1 x}\phi^{N}_1
+ t^{hh}_{11sh} e^{ik_2 x}\phi^{N}_2
+ t^{eh}_{12sh} e^{-ik_1 x}\phi^{N}_3
+ t^{hh}_{12sh} e^{ik_2 x}\phi^{N}_4, 
&& x<0, \\[6pt]
\psi_{IP}(x) &=
r^{eh}_{11sh} e^{iq_{1e} x}\phi^{S}_1
+ \left(e^{iq_{1h} x} +r^{hh}_{11sh} e^{-iq_{1h} x}\right)\phi^{S}_2
+ r^{eh}_{12sh} e^{iq_{2e} x}\phi^{S}_3
+ r^{hh}_{12sh} e^{-iq_{2h} x}\phi^{S}_4, 
&& x>0.
\end{align}
\end{subequations}
We can calculate the scattering amplitudes using the same boundary condition Eq.~\ref{bcapp} and the current conservation. The scattering amplitudes are
$a_{11sh}=r^{eh}_{11sh},\;
a_{12sh}=\sqrt{\tfrac{|u_2|^2-|v_2|^2}{|u_1|^2-|v_1|^2}}\,r^{eh}_{12sh},\;
b_{11sh}=r^{hh}_{11sh},\;
b_{12sh}=\sqrt{\tfrac{|u_2|^2-|v_2|^2}{|u_1|^2-|v_1|^2}}\,r^{hh}_{12sh},\;
c_{11sh}=\tfrac{1}{\sqrt{|u_1|^2-|v_1|^2}}\,t^{hh}_{11sh},\;
c_{12sh}=\tfrac{1}{\sqrt{|u_1|^2-|v_1|^2}}\,t^{hh}_{12sh},\;
d_{11sh}=\tfrac{1}{\sqrt{|u_1|^2-|v_1|^2}}\,t^{eh}_{11sh},\;
d_{12sh}=\tfrac{1}{\sqrt{|u_2|^2-|v_2|^2}}\,t^{eh}_{12sh}.$
The corresponding probabilities are
$A_{11sh}=|a_{11sh}|^2,\;
A_{12sh}=|a_{12sh}|^2,\;
B_{11sh}=|b_{11sh}|^2,\;
B_{12sh}=|b_{12sh}|^2,\;
C_{11sh}=|c_{11sh}|^2,\;
C_{12sh}=|c_{12sh}|^2,\;
D_{11sh}=|d_{11sh}|^2,\;
D_{12sh}=|d_{12sh}|^2.$

% Amplitudes:
% \begin{align*}
% a_{11sh} &= r^{eh}_{11sh}, \quad
% &a_{12sh} &= \sqrt{\frac{|u_2|^2 - |v_2|^2}{|u_1|^2 - |v_1|^2}} \, r^{eh}_{12sh}, \\[1mm]
% b_{11sh} &= r^{hh}_{11sh}, \quad
% &b_{12sh} &= \sqrt{\frac{|u_2|^2 - |v_2|^2}{|u_1|^2 - |v_1|^2}} \, r^{hh}_{12sh}, \\[1mm]
% c_{11sh} &= \frac{1}{\sqrt{|u_1|^2 - |v_1|^2}} \, t^{hh}_{11sh}, \quad
% &c_{12sh} &= \frac{1}{\sqrt{|u_1|^2 - |v_1|^2}} \, t^{hh}_{12sh}, \\[1mm]
% d_{11sh} &= \frac{1}{\sqrt{|u_1|^2 - |v_1|^2}} \, t^{eh}_{11sh}, \quad
% &d_{12sh} &= \frac{1}{\sqrt{|u_2|^2 - |v_2|^2}} \, t^{eh}_{12sh}.
% \end{align*}

% Probabilities:
% \[
% A_{11sh}=|a_{11sh}|^2,\;\; A_{12sh}=|a_{12sh}|^2,\;\;
% B_{11sh}=|b_{11sh}|^2,\;\; B_{12sh}=|b_{12sh}|^2,\;\;
% C_{11sh}=|c_{11sh}|^2,\;\; C_{12sh}=|c_{12sh}|^2,\;\;
% D_{11sh}=|d_{11sh}|^2,\;\; D_{12sh}=|d_{12sh}|^2.
% \]

% ===============================
% Case 4: Hole incident on band 2
% ===============================
The wave function for a hole incident on band 2 is
\begin{subequations}
\begin{align}
\psi_{N}(x) &=
 t^{eh}_{21sh} e^{-ik_1 x}\phi^{N}_1
+ t^{hh}_{21sh} e^{ik_2 x}\phi^{N}_2
+ t^{eh}_{22sh} e^{-ik_1 x}\phi^{N}_3
+ t^{hh}_{22sh} e^{ik_2 x}\phi^{N}_4, 
&& x<0, \\[6pt]
\psi_{IP}(x) &=
r^{eh}_{21sh} e^{iq_{1e} x}\phi^{S}_1
+ r^{hh}_{21sh} e^{-iq_{1h} x}\phi^{S}_2
+ r^{eh}_{22sh} e^{iq_{2e} x}\phi^{S}_3
+\left(e^{iq_{2h} x} + r^{hh}_{22sh} e^{-iq_{2h} x}\right)\phi^{S}_4, 
&& x>0.
\end{align}
\end{subequations}

We can calculate the scattering amplitudes using the same boundary condition Eq.~\ref{bcapp} and the current conservation.
The scattering amplitudes are
$a_{21sh}=\sqrt{\tfrac{|u_1|^2-|v_1|^2}{|u_2|^2-|v_2|^2}}\,r^{eh}_{21sh},\;
a_{22sh}=r^{eh}_{22sh},\;
b_{21sh}=\sqrt{\tfrac{|u_1|^2-|v_1|^2}{|u_2|^2-|v_2|^2}}\,r^{hh}_{21sh},\;
b_{22sh}=r^{hh}_{22sh},\;
c_{21sh}=\tfrac{1}{\sqrt{|u_2|^2-|v_2|^2}}\,t^{hh}_{21sh},\;
c_{22sh}=\tfrac{1}{\sqrt{|u_2|^2-|v_2|^2}}\,t^{hh}_{22sh},\;
d_{21sh}=\tfrac{1}{\sqrt{|u_2|^2-|v_2|^2}}\,t^{eh}_{21sh},\;
d_{22sh}=\tfrac{1}{\sqrt{|u_2|^2-|v_2|^2}}\,t^{eh}_{22sh}.$
The corresponding probabilities are
$A_{21sh}=|a_{21sh}|^2,\;
A_{22sh}=|a_{22sh}|^2,\;
B_{21sh}=|b_{21sh}|^2,\;
B_{22sh}=|b_{22sh}|^2,\;
C_{21sh}=|c_{21sh}|^2,\;
C_{22sh}=|c_{22sh}|^2,\;
D_{21sh}=|d_{21sh}|^2,\;
D_{22sh}=|d_{22sh}|^2.$

% Amplitudes:
% \begin{align*}
% a_{21sh} &= \sqrt{\frac{|u_1|^2 - |v_1|^2}{|u_2|^2 - |v_2|^2}} \, r^{eh}_{21sh}, \quad
% &a_{22sh} &= r^{eh}_{22sh}, \\[1mm]
% b_{21sh} &= \sqrt{\frac{|u_1|^2 - |v_1|^2}{|u_2|^2 - |v_2|^2}} \, r^{hh}_{21sh}, \quad
% &b_{22sh} &= r^{hh}_{22sh}, \\[1mm]
% c_{21sh} &= \frac{1}{\sqrt{|u_2|^2 - |v_2|^2}} \, t^{hh}_{21sh}, \quad
% &c_{22sh} &= \frac{1}{\sqrt{|u_2|^2 - |v_2|^2}} \, t^{hh}_{22sh}, \\[1mm]
% d_{21sh} &= \frac{1}{\sqrt{|u_2|^2 - |v_2|^2}} \, t^{eh}_{21sh}, \quad
% &d_{22sh} &= \frac{1}{\sqrt{|u_2|^2 - |v_2|^2}} \, t^{eh}_{22sh}.
% \end{align*}

% Probabilities:
% \[
% A_{21sh}=|a_{21sh}|^2,\;\; A_{22sh}=|a_{22sh}|^2,\;\;
% B_{21sh}=|b_{21sh}|^2,\;\; B_{22sh}=|b_{22sh}|^2,\;\;
% C_{21sh}=|c_{21sh}|^2,\;\; C_{22sh}=|c_{22sh}|^2,\;\;
% D_{21sh}=|d_{21sh}|^2,\;\; D_{22sh}=|d_{22sh}|^2.
% \]

Below we discuss the scattering matrix, 
$C_{\text{out}} = (c_{e1}^-, c_{e2}^-, c_{h1}^-, c_{h2}^-, b_{e1}^-, b_{e2}^-, b_{h1}^-, b_{h2}^-)^{T},\;
C_{\text{in}} = (c_{e1}^+, c_{e2}^+, c_{h1}^+, c_{h2}^+, b_{e1}^+, b_{e2}^+, b_{h1}^+, b_{h2}^+)^{T},\;
C_{\text{out}} = S\,C_{\text{in}}.$

\[S=
\begin{pmatrix}
s_{11}^{ee11} & s_{11}^{ee12} & s_{11}^{eh11} & s_{11}^{eh12} &
s_{12}^{ee11} & s_{12}^{ee12} & s_{12}^{eh11} & s_{12}^{eh12} \\
s_{11}^{ee21} & s_{11}^{ee22} & s_{11}^{eh21} & s_{11}^{eh22} &
s_{12}^{ee21} & s_{12}^{ee22} & s_{12}^{eh21} & s_{12}^{eh22} \\
s_{11}^{he11} & s_{11}^{he12} & s_{11}^{hh11} & s_{11}^{hh12} &
s_{12}^{he11} & s_{12}^{he12} & s_{12}^{hh11} & s_{12}^{hh12} \\
s_{11}^{he21} & s_{11}^{he22} & s_{11}^{hh21} & s_{11}^{hh22} &
s_{12}^{he21} & s_{12}^{he22} & s_{12}^{hh21} & s_{12}^{hh22} \\
s_{21}^{ee11} & s_{21}^{ee12} & s_{21}^{eh11} & s_{21}^{eh12} &
s_{22}^{ee11} & s_{22}^{ee12} & s_{22}^{eh11} & s_{22}^{eh12} \\
s_{21}^{ee21} & s_{21}^{ee22} & s_{21}^{eh21} & s_{21}^{eh22} &
s_{22}^{ee21} & s_{22}^{ee22} & s_{22}^{eh21} & s_{22}^{eh22} \\
s_{21}^{he11} & s_{21}^{he12} & s_{21}^{hh11} & s_{21}^{hh12} &
s_{22}^{he11} & s_{22}^{he12} & s_{22}^{hh11} & s_{22}^{hh12} \\
s_{21}^{he21} & s_{21}^{he22} & s_{21}^{hh21} & s_{21}^{hh22} &
s_{22}^{he21} & s_{22}^{he22} & s_{22}^{hh21} & s_{22}^{hh22} \\
\end{pmatrix}
\]

The $S-$matrix is unitary, i.e., $S^\dagger S=I_8$ ($I_8$ is the $8\times8$ identity matrix).

\section{Calculation of finite temperature quantum noise($Q_{11}$)}
\label{quantum noise appendix}

% \subsection{Quantum noise}
% \label{quantum noise theory}
Quantum noise auto- and cross-correlations describe the fluctuations in charge current between terminals in mesoscopic systems such as the NIN junction. Quantum noise can also be investigated in hybrid junctions such as N-I-IP junctions. In such junctions, the current-current correlations between terminals $i$ and $j$ capture the interplay of electron and hole transport($i$, and $j$ stand for left normal metal and Iron Pnictide superconductor, respectively). The quantum noise correlation function can be defined as
\begin{equation}
Q^{xy}_{ij}(t - t') = \langle \Delta I^x_i(t) \Delta I^y_j(t') + \Delta I^y_j(t') \Delta I^x_i(t) \rangle,
\label{timefluctuaions}
\end{equation}
where $\Delta I^x_i(t) = I^x_i(t) - \langle I^x_i(t) \rangle$ denotes the fluctuation in current of type $x$ (with $x \in \{e, h\}$) and $i,j\in \{1,2\}$ denotes terminals.

The Fourier transform of the above Eq. \ref{timefluctuaions} gives frequency-dependent fluctuations,
\begin{equation}
\delta(\omega + \bar{\omega}) Q^{xy}_{NP}(\omega) = \frac{1}{2\pi} \langle \Delta I^x_N(\omega) \Delta I^y_P(\bar{\omega}) + \Delta I^y_P(\bar{\omega}) \Delta I^x_N(\omega) \rangle.
\end{equation}
In particular, for a two-terminal N-I-IP setup, the zero-frequency limit of the auto-correlation function is defined as in Refs.~\cite{datta1996}. Unlike finite-frequency noise, zero-frequency shot noise has a simpler theoretical interpretation, is less affected by environmental factors, and requires less complex experimental setups. It directly reveals key transport properties such as effective charge and transmission statistics, making it a robust and practical diagnostic tool in condensed matter physics and mesoscopic systems~\cite{Blanter2000},

\begin{align}
{Q^{mn}_{\alpha \beta,i j} (\omega =0)} &=  \frac{e^2}{h} 
\int \sum_{\substack{k,l \in \{1, 2\}, \\ \gamma,\delta \in \{e,h\}\\ m_1,n_1 \in \{1,2\}}}
sgn(\alpha) sgn(\beta) A_{k,\gamma;l,\delta}^{m_1n_1}(i,  \alpha,E) A_{l,\delta ;k,\gamma}^{n_1m_1}(j , \beta,E)   \times [ \textit{f}_{k \gamma}(E) (1-\textit{f}_{l \delta}(E)) +  \textit{f}_{l \delta}(E) (1-\textit{f}_{k \gamma}(E))] dE, 
\label{eqn:Qnoise}
\end{align}
where $A_{k,\gamma;l,\delta}^{m_1n_1}(i,  \alpha,E)=\delta_{ik}\delta_{il}\delta_{\alpha\gamma}\delta_{\alpha \delta}\delta_{m_1n_1}-s_{ik}^{\alpha\gamma mm_1^\dagger} s_{il}^{\alpha\delta mn_1}$, 
$Q_{ee,11} = Q_{ee,11}^{11} + Q_{ee,11}^{12} + Q_{ee,11}^{21} + Q_{ee,11}^{22},\quad
Q_{eh,11} = Q_{eh,11}^{11} + Q_{eh,11}^{12} + Q_{eh,11}^{21} + Q_{eh,11}^{22},
Q_{he,11} = Q_{he,11}^{11} + Q_{he,11}^{12} + Q_{he,11}^{21} + Q_{he,11}^{22},\quad
Q_{hh,11} = Q_{hh,11}^{11} + Q_{hh,11}^{12} + Q_{hh,11}^{21} + Q_{hh,11}^{22}.$
\\
The finite temperature quantum noise is, 
$Q_{11} = Q_{ee,11}+Q_{eh,11}+Q_{he,11}+Q_{hh,11}$. The following equations are used from the unitarity of the scattering matrix to obtain the quantum noise.
% \vspace{-0.5cm}

\begin{equation}
\begin{aligned}
& A_{1n} + A_{21nh} + B_{1n} + B_{21n} + C_{1s} + C_{21s}
+ D_{11s} + D_{21s} 
&&= 1 \\[4pt]
& A_{22nh} + A_{2n} + B_{22n} + B_{12n} + C_{22s} + C_{2s}
+ D_{22s} + D_{12s} 
&&= 1 \\[4pt]
& A_{1n} + A_{21n} + B_{1h} + B_{21h} + C_{1sh} + C_{21sh}
+ D_{11s} + D_{21s} 
&&= 1 \\[4pt]
& A_{22nh} + A_{2n} + B_{22nh} + B_{12h} + C_{22sh} + C_{2sh}
+ D_{22s} + D_{12s} 
&&= 1 \\[6pt]
& b_{n}a_{1n}^* + b_{21n}a_{21n}^* + a_{1h}b_{1h}^*
+ a_{21nh}b_{21nh}^* + d_{21sh}c_{21sh}^*
+ d_{12sh}c_{12sh}^* + c_{1s}c_{21s}^*
+ c_{2s}c_{22s}^* 
&&= 0 \\[4pt]
& b_{21n}a_{2n}^* + b_{22n}a_{22nh}^* + a_{21h}b_{12h}^*
+ a_{22nh}b_{22nh}^* + d_{21sh}c_{11s}^*
+ d_{12sh}c_{21s}^* + c_{2s}c_{11s}^*
+ c_{22s}c_{21s}^* 
&&= 0 \\[4pt]
& a_{21n}a_{22n}^* + a_{1n}a_{2n}^* + b_{21nh}b_{22nh}^*
+ b_{1h}b_{12h}^* + c_{21sh}c_{22sh}^*
+ c_{1sh}c_{2sh}^* + d_{11s}d_{22s}^* 
&&= 0
\end{aligned}
\end{equation}

The quantum noise is the sum of quantum shot noise and quantum {equilibrium noise}, i.e.,
% \vspace{-1.6cm}
\begin{align}
    % Q_{11}=&Q_{11}^{sh}+Q_{11}^{th}\\
    \quad Q_{11}=&\frac{2e^2}{h}\int_0^\infty dE[c_1(f_{1e}-f_{1h})^2+c_2(f_{1e}-f_{2e})^2+c_3(f_{1h}-f_{2e})^2 +
t_{1}\,f_{1h}(1-f_{1h})
+ t_{2}\,f_{1e}(1-f_{1e})
+ t_{3}\,f_{2e}(1-f_{2e})
\Big],
\label{qnapp}
\end{align}

% \begin{align}
%     Q_{11}^{sh}=\frac{2e^2}{h}\int_0^\infty dE[c_1(f_{1e}-f_{1h})^2+c_2(f_{1e}-f_{2e})^2+c_3(f_{1h}-f_{2e})^2],
% \label{qshapp}
% \end{align}
\begin{align}
c_1 &= \Big[
 (A_{11n}+A_{21n})(B_{11nh}+B_{21nh})
 + (A_{11nh}+A_{21nh})(B_{11n}+B_{21n}) 
 + (A_{22nh}+A_{12nh})(B_{22n}+B_{12n})
 + (A_{22n}+A_{12n})(B_{22nh}+B_{12nh})
 \Big] \nonumber \\
&\quad - 2\,\mathrm{Re}\Big[
 (b_{11n} a_{11n}^* + b_{21n} a_{21n}^*)
 (b_{11n}^* a_{11n} + b_{21n}^* a_{21n})
 + (b_{21n} a_{22n}^* + b_{11n} a_{12n}^*)
 (b_{12n}^* a_{11n} + b_{22n}^* a_{21n}) \nonumber \\
&\qquad + (b_{21n} a_{22n}^* 
 + b_{11n} a_{12n}^*)
 (b_{12n}^* a_{11n} + b_{22n}^* a_{21n})
 + (b_{22n}^* a_{22n} + b_{12n}^* a_{12n})
 (b_{12n} a_{12n}^* + b_{22n} a_{22n}^*)
 \Big]  \nonumber \\
&\qquad + 2\,\mathrm{Re}\Big[
 (b_{12n}^* b_{11n} + b_{22n}^* b_{21n})
 (a_{21n} a_{22n}^* + a_{11n} a_{12n}^*)
 + (b_{12n} b_{11n}^* + b_{22n} b_{21n}^*)
 (a_{21n}^* a_{22n} + a_{11n}^* a_{12n})
 \Big], \nonumber \\[8pt]
c_2 &= \Big[
 (B_{11n}+B_{21n})(1-A_{11nh}-A_{21nh}-B_{11n}-B_{21n})
 + (A_{11n}+A_{21n})(1-A_{11n}-A_{21n}-B_{11nh}-B_{21nh}) \nonumber \\
&\qquad
 + (B_{22n}+B_{12n})(1-A_{22nh}-A_{12nh}-B_{22n}-B_{12n}) 
 + (A_{22n}+A_{12n})(1-A_{22n}-A_{12n}-B_{22nh}-B_{12nh})
 \Big] \nonumber \\
&\quad + 2\,\mathrm{Re}\Big[
 (b_{11n} a_{11n}^* + b_{21n} a_{21n}^*)
 (b_{11n}^* a_{11n} + b_{21n}^* a_{21n}
 + a_{11nh}^* b_{11nh} + a_{21nh}^* b_{21nh}) \nonumber \\
&\qquad
 + (b_{21n} a_{22n}^* + b_{11n} a_{12n}^*)
 (b_{21n}^* a_{22n} + b_{11n}^* a_{12n}
 + a_{22nh}^* b_{22nh} + a_{12nh}^* b_{12nh}) \nonumber \\
&\qquad
 + (b_{12n} a_{21n}^* + b_{22n} a_{22n}^*)
 (b_{12n}^* a_{21n} + b_{22n}^* a_{22n}
 + a_{21nh}^* b_{21nh} + a_{22nh}^* b_{22nh}) \nonumber \\
&\qquad
 + (b_{12n} a_{12n}^* + b_{22n} a_{22n}^*)
 (b_{12n}^* a_{12n} + b_{22n}^* a_{22n}
 + a_{12nh}^* b_{12nh} + a_{22nh}^* b_{22nh})
 \Big] \nonumber \\
&\quad - 2\,\mathrm{Re}\Big[
 (b_{12n} b_{22n}^* + b_{11n} b_{21n}^*)
 (a_{21n} a_{22n}^* + a_{11n} a_{12n}^*
 + b_{21nh}^* b_{22nh} + b_{11nh}^* b_{12nh}) \nonumber \\
&\qquad
 + (a_{21n} a_{22n}^* + a_{11n} a_{12n}^*)
 (a_{21n}^* a_{22n} + a_{11n}^* a_{12n}
 + b_{21nh} b_{22nh} + b_{11nh} b_{12nh})
 \Big], \nonumber \\[8pt]
c_3 &= \Big[
 (A_{11nh}+A_{21nh})(1-A_{11n}-A_{21n}-B_{11n}-B_{21n})
 + (B_{11nh}+B_{21nh})(1-A_{11n}-A_{21n}-B_{11nh}-B_{21nh}) \nonumber \\
&\qquad
 + (A_{22nh}+A_{12nh})(1-A_{22n}-A_{12n}-B_{22n}-B_{12n})
 + (B_{22nh}+B_{12nh})(1-A_{22n}-A_{12n}-B_{22nh}-B_{12nh})
 \Big] \nonumber \\
&\quad + 2\,\mathrm{Re}\Big[
 (a_{11nh} b_{11nh}^* + a_{21nh} b_{21nh}^*)
 (b_{11n}^* a_{11n} + b_{21n}^* a_{21n}
 + a_{11nh}^* b_{11nh} + a_{21nh}^* b_{21nh}) \nonumber \\
&\qquad
 + (a_{11nh} b_{12nh}^* + a_{21nh} b_{22nh}^*)
 (b_{21n}^* a_{22n} + b_{11n}^* a_{12n}
 + a_{22nh}^* b_{22nh} + a_{12nh}^* b_{12nh}) \nonumber \\
&\qquad
 + (a_{22nh} b_{21nh}^* + a_{12nh} b_{11nh}^*)
 (b_{12n}^* a_{21n} + b_{22n}^* a_{22n}
 + a_{12nh}^* b_{21nh} + a_{22nh}^* b_{12nh}) \nonumber \\
&\qquad
 + (a_{22nh} b_{22nh}^* + a_{12nh} b_{12nh}^*)
 (b_{22n}^* a_{22n} + b_{12n}^* a_{12n}
 + a_{22nh}^* b_{22nh} + a_{12nh}^* b_{12nh})
 \Big] \nonumber \\
&\quad - 2\,\mathrm{Re}\Big[
 (a_{11nh} a_{21nh}^* + a_{22nh} a_{12nh}^*)
 (a_{21n} a_{22n}^* + a_{11n} a_{12n}^*
 + b_{21nh}^* b_{22nh} + b_{11nh}^* b_{12nh}) \nonumber \\
&\qquad
 + (b_{12nh} b_{22nh}^* + b_{11nh} b_{21nh}^*)
 (a_{12n}^* a_{22n} + a_{11n}^* a_{21n}
 + b_{12n}^* b_{22n} + b_{11n}^* b_{21n})
 \Big]. \nonumber
\label{c1c2c3}
\end{align}
\begin{figure}[H]
    \centering
    \includegraphics[width=1.0\linewidth]{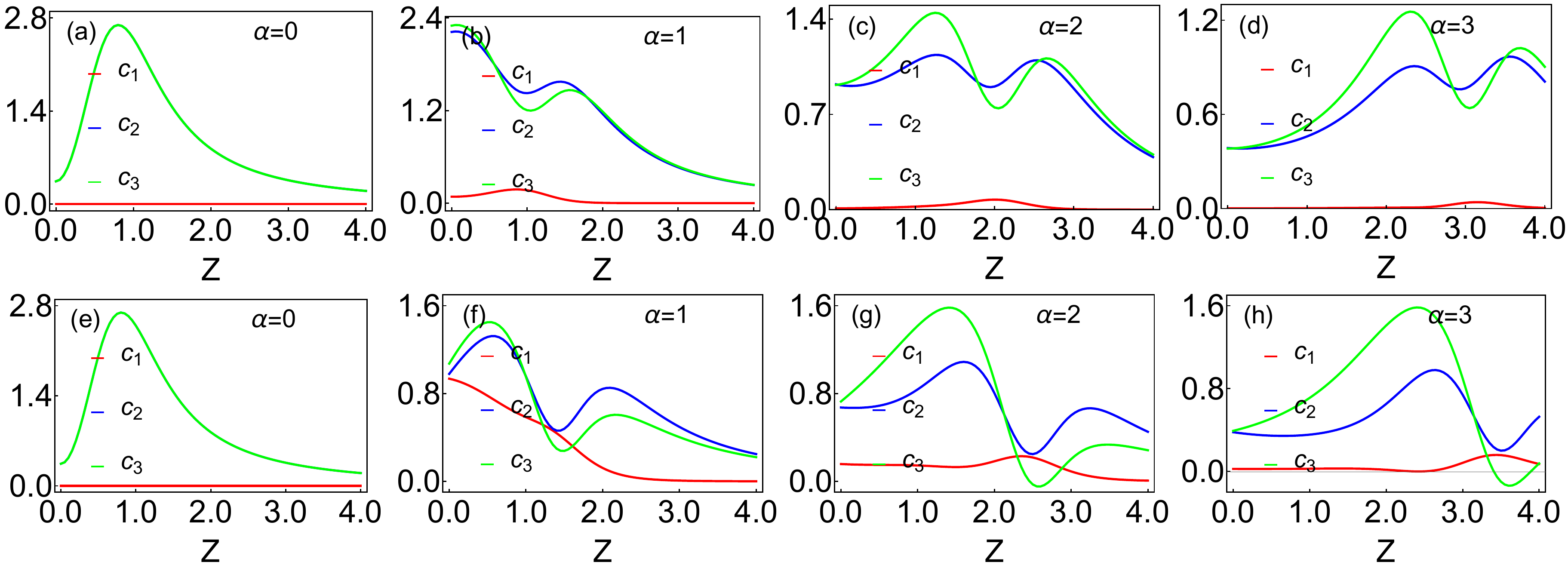}
\caption{Plot $c_1,c_2$ and $c_3$ vs brrier strength($Z$) at different interband coupling strengths. (a--d) is for $S_{++}$ pairing symmery, (e--h) for $S_{+-}$ pairing symmetry. (a, e) $\alpha=0$, (b, f) $\alpha=1$, (c, g) $\alpha=2$ and (d, h) $\alpha=3$}

\label{noiseapp}
\end{figure}

\begin{align*}
t_1 =& -2 \Big[
2 \operatorname{Re}\!\Big((a_{21nh} a_{22nh} a_{11nh}^* a_{12nh}^*)
            (b_{11n} b_{21nh}^*+b_{12nh} b_{22nh}^*)\Big)  + A_{11nh}(B_{11nh}+B_{12nh}-1)
  + A_{21nh}(B_{21nh}+B_{22nh}-1)  \\
& + A_{22nh}(B_{21nh}+B_{22nh}-1)
  + A_{12nh}(B_{11nh}+B_{12nh}-1)
\Big]  + 2 \Big[
2 \operatorname{Re}\!\big(a_{21nh} a_{12nh} a_{11nh}^* a_{22nh}^*
           + b_{21nh} b_{12nh} b_{11nh}^* b_{22nh}^*\big)  \\
& + B_{21nh}(B_{22nh}-1)
- B_{22nh}-B_{12nh}+1
\Big]  + (A_{11nh}+A_{21nh})^2
+ 2A_{11nh}A_{12nh}
+ 2A_{21nh}A_{22nh}  \\
& + A_{22nh}^2
+ 2A_{22nh}A_{12nh}
+ A_{12nh}^2   + B_{11nh}^2
+ 2B_{11nh}(B_{21nh}+B_{12nh}-1)
+ B_{21nh}^2  + (B_{22nh}+B_{12nh})^2
+ (c_1 + c_3) .\\
t_2 =& -2 \Big[ 
2 \operatorname{Re}\!\Big((a_{11n} a_{21n}^*+a_{12n} a_{22n}^*)
            (b_{21n} b_{11n}^*+b_{22n} b_{12n}^*)\Big) \\
& + A_{11n}(B_{11n}+B_{12n}-1)
  + A_{21n}(B_{21n}+B_{22n}-1) + A_{22n}(B_{21n}+B_{22n}-1)
  + A_{12n}(B_{11n}+B_{12n}-1)
\Big]  \\
& + 2 \Big[
2 \operatorname{Re}\!\big(a_{21n} a_{12n} a_{11n}^* a_{22n}^*
           + b_{21n} b_{12n} b_{11n}^* b_{22n}^*\big)
+ A_{21n}A_{22n}+A_{22n}A_{12n}+1
\Big]   + (A_{11n}+A_{21n})^2
+ 2A_{11n}A_{12n}
+ A_{22n}^2 + A_{12n}^2  \\
& + B_{11n}^2
+ 2B_{11n}(B_{21n}+B_{12n}-1)
+ B_{21n}^2  + 2B_{21n}(B_{22n}-1)
+ (B_{22n}+B_{12n})^2
-2B_{22n}-2B_{12n}
+ c_1 + c_2.\\
t_3 =& -A_{11nh}-A_{11n}-A_{21n}-A_{21nh}-A_{22n}-A_{22nh} -A_{12nh}-A_{12n}-B_{11nh}-B_{11n}-B_{21n}-B_{21nh}  \\
      & -B_{22n}-B_{22nh}-B_{12nh}-B_{12n}+4 .
\end{align*}

 \end{widetext}

\clearpage
\bibliography{cite.bib}
\end{document}